\documentclass{article}

\PassOptionsToPackage{numbers, compress}{natbib}
\usepackage[dblblindworkshop, final]{neurips_2025}
\usepackage[utf8]{inputenc} % allow utf-8 input
\usepackage[T1]{fontenc}    % use 8-bit T1 fonts
\usepackage{hyperref}       % hyperlinks
\usepackage{url}            % simple URL typesetting
\usepackage{booktabs}       % professional-quality tables
\usepackage{amsfonts}       % blackboard math symbols
\usepackage{nicefrac}       % compact symbols for 1/2, etc.
\usepackage{microtype}      % microtypography
\usepackage{xcolor}         % colors
\usepackage{hyperref}
\usepackage{booktabs}
\usepackage{graphicx}
\usepackage{caption}
\usepackage{subcaption}
\usepackage[inkscapelatex=false]{svg}
\usepackage{multirow}
\usepackage{enumitem}
\usepackage{dsfont}
\usepackage{amsmath}
\usepackage{url}
\newcommand{\triple}{(BCC/B2/B2 volume \%) }

\title{Preference Learning from Physics-Based Feedback: Tuning Language Models to Design BCC/B2 Superalloys}
\workshoptitle{AI4Mat: AI for Accelerated Materials Design}
\author{
Satanu Ghosh$^{1}$ \quad Collin Holgate$^{2}$ \quad Neal R. Brodnik$^{2}$ \quad Doug Downey$^{4}$ \\
\bf Samantha Daly$^{3}$ \quad Tresa M. Pollock$^{2}$  \quad \bf Samuel Carton$^{1}$  \\
$^{1}$Department of Computer Science, University of New Hampshire \\ 
$^{2}$Materials Department, University of California, Santa Barbara \\
$^{3}$Department of Mechanical Engineering, University of California, Santa Barbara \\
$^{4}$Allen Institute for Artificial Intelligence\\
\texttt{\{satanu.ghosh, samuel.carton\}@unh.edu} \\
\texttt{\{holgate, nrbodnik, tresap, samdaly\}@ucsb.edu} \\
\texttt{\{dougd\}@allenai.org}}

\begin{document}

\maketitle

\begin{abstract}
We apply preference learning to the task of language model-guided design of novel structural alloys. In contrast to prior work that focuses on generating stable inorganic crystals, our approach targets the synthesizeability of a specific structural class: BCC/B2 superalloys, an underexplored family of materials with potential applications in extreme environments. Using three open-weight models (LLaMA-3.1, Gemma-2, and OLMo-2), we demonstrate that language models can be optimized for multiple design objectives using a single, unified reward signal through Direct Preference Optimization (DPO). Unlike prior approaches that rely on heuristic or human-in-the-loop feedback (costly), our reward signal is derived from thermodynamic phase calculations, offering a scientifically grounded criterion for model tuning. To our knowledge, this is the first demonstration of preference-tuning a language model using physics-grounded feedback for structural alloy design. The resulting framework is general and extensible, providing a path forward for intelligent design-space exploration across a range of physical science domains.
\end{abstract}

\section{Introduction}
% Language models (LMs) have been increasingly studied as a 

% Recent work has shown that supervised fine-tuning (SFT) can induce language models (LMs) to capture generalized notions of thermodynamic stability and generate novel, stable material compositions \cite{gruver2024fine, antunes_crystal_2024}. However, while this approach is appropriate for producing a baseline LM that accedes to a  binary, (dis)qualifying property like stability, it doesn't offer a way to optimize the LM toward desired continuous-valued properties (e.g. toughness, ductility) in its generations. For this next step, reinforcement learning-based preference tuning methods such as proximal policy optimization (PPO) \cite{schulman_proximal_2017} and direct policy optimization (DPO) \cite{rafailov2023direct} are more suitable. 

% In this paper we apply preference tuning for the first time (that we know of) to the task of optimizing the properties of LM-generated materials based on feedback from physics-based simulations. Specifically, we apply DPO to optimize the generation of viable BCC/B2 ``superalloys'' based on feedback from Thermo-Calc, a popular CALPHAD thermodynamic simulation tool. 

Materials discovery is
% is crucially important to numerous industries, but 
challenging because of large design spaces sparsely covered by empirical results, and the intrinsic nonlinearity and multiobjectivity of materials design problems. Computational materials science addresses this sparsity by modeling from simulations, often based on density functional theory (DFT)~\cite{kohn1996density}, and knowledge bases such as the Inorganic Crystal Structure Database (ICSD)~\cite{Zagorac:in5024}. When trained on these sources, discriminative machine learning models can cheaply predict properties of unknown materials (forward design), while generative models can propose materials with favorable properties (inverse design). 

Large language models (LMs), when trained or prompted appropriately, can generate descriptions of new materials. They are held as a potential accelerant to material discovery for their ability to draw on parametrically-encoded and retrieved domain knowledge to propose materials more likely to have desirable properties \cite{li_materials_2025,brodnik2023perspective}. 
Prior work on LM-driven inverse design mostly falls into two categories. The first trains smaller local LMs, mostly via supervised fine-tuning (SFT) to generate candidate materials satisfying a single basic criterion, commonly thermodynamic stability~\cite{gruver2024fine,sriram2024flowllm,antunes_crystal_2024}. The second category involves using a larger API-based LM as part of a search/optimization procedure to identify high-quality outputs according to multi-objective criteria, often in an multi-agent setup (e.g. \cite{gan2025large, yang2024generative, lai2025prim}). 

In this paper, we explore an intermediate step: using preference tuning to align local language models toward more optimal arbitrary downstream property values. Specifically, we use offline preference learning based on multiobjective feedback from a physical simulation model to nudge the LM into a ``high-reward'' output space where its generations are more likely to be of high quality while still remaining diverse within the chosen design space.

We apply this approach to the task of structural alloy design, specifically BCC/B2 ``superalloys'' consisting of a matrix of disordered, body-centered cubic (BCC) material surrounding precipitates of ordered BCC (B2) material. This type of alloy, consisting of two distinct phases, is a promising recent direction in extreme-environment structural alloys. By adding a second phase, they potentially address the structural weakness that existing alloys tend to exhibit at high temperatures (>1000°C) \cite{Kube2024BCCB2, precipitates_wang2018coherent, precipitates_yurchenko2021refractory}. However, inducing the stable formation of two complementary phases is nontrivial. Any generative modeling approach needs to produce candidates that are both practically viable as well as potentially useful. 
Our approach generates superalloy candidates in the form of a composition for the BCC matrix, the B2 precipitate, and a suggested volume percentage for the B2. We apply a two-step modeling process mirroring conventional LM preference alignment. Starting with a known set of BCC and B2 compositions, we apply supervised fine-tuning (SFT) to three local instruction tuned language models (LLaMA 3.1 8B, Gemma-2-9B OLMo-2-7B) to produce 
% \{BCC/B2/B2 volume percentage\} 
\triple triples. 
We then use feedback on generated candidates from Thermo-Calc~\cite{andersson2002thermocalc}, a popular thermodynamic simulation tool, to produce a multiobjective reward score for each candidate based on expert-designed heuristics. Finally, we use these scores for direct preference optimization (DPO), to push the models into a higher-reward output mode.

In our evaluation, we demonstrate that our SFT-tuned models are capable of generating valid alloy compositions that uniformly span the design space and exhibit novelty with respect to both the training data and existing entries in the Materials Project database. We further show that the DPO-tuned models, with the exception of OLMo, demonstrate improved average reward scores while retaining a high degree of diversity in their outputs. Our findings indicate that local language models can be effectively optimized for multiple design objectives using a single, unified reward signal.
By comparison, larger state-of-the-art API-based LMs are able to suggest high-reward alloy compositions without tuning, but tend to hyper-fixate on specific elements and combinations, leading to limited exploration of the specified design space, a behavior resistant to prompt engineering. We conclude by outlining key takeaways and discussing how this preference tuning framework can potentially be extended to future materials discovery tasks and other domains within the physical sciences.

% LMs tuned to produce material structure can play a role either in agentic setups (e.g. \cite{lai2025prim} as a candidate generator, or in more traditional search algorithms such as Bayesian Optimization (e.g. \cite{hastings_accelerated_2025, wang_bayesian_2022}). In either case, while it is important to for generated materials to be valid (i.e. stable), it is also important for them to be diverse, while tilting them to

%  Need some content that talks about what is special about our approach
In summary, our contributions are as follows:
\begin{enumerate}[noitemsep,nolistsep,topsep=0pt,leftmargin=20pt]
	\item To our knowledge, this work presents the first instances of:
	\begin{itemize}[noitemsep,nolistsep,topsep=0pt,leftmargin=10pt]
		\item Preference tuning for language models in the context of materials composition generation.
		\item Guiding a language model to generate materials compositions aligned with a multi-objective design goal, moving beyond optimization for a single figure of merit (e.g., thermodynamic stability).
	\end{itemize}
	\item We propose a general and extensible framework for scientist-informed candidate generation in non-parametric design spaces, leveraging offline feedback from physics-based simulations.
	\item We apply our framework to a real-world challenge in materials design—specifically, the discovery of BCC/B2 superalloys, moving away from general-purpose stable crystal generation toward targeted, high-impact alloy design. 
\end{enumerate}

Code and data are available at:\\
\url{https://github.com/SatanuG/BCC-B2-Super-Alloy-Generator}.

\section{Related Work}

\paragraph{Conventional superalloy discovery}

% \sam{What are the conventional approaches to computational alloy discovery?  Tresa: changed the first 2 sentences}

% Historically, design and development of new alloys has been an extraordinarily slow process, often requiring more than a decade, due to complex iterative experimental loops.  
% Within the past decade, advances in computation has enabled materials discovery via \textit{ab-initio} simulation methods, such as density functional theory \cite{kohn1996density} and molecular dynamics, which use foundational physics to model material interactions at the scale of individual atoms \cite{DFT_bartolotti1996introduction, DFT_geerlings2003conceptual, Visual_MD_humphrey1996vmd, VASP_kresse1996software, VASP_kresse1996efficiency, VASP_kresse1993ab}.

Superalloys are a class of multiphase alloys that combine a ductile matrix phase with high-strength precipitates to produce a material that is both strong and tough at elevated temperatures. Current commercial superalloys, such as the Inconel and Ren\'e classes of alloys, have a face-centered-cubic (FCC) matrix and $\mathrm{L1_2}$ intermetallic precipitates.  However, modern operation demands have now extended to temperatures beyond the design limit of any known FCC/$\mathrm{L1_2}$ superalloy.  In the search for even higher temperature alloys, significant interest has been directed at systems composed of a body-centered-cubic (BCC) matrix with ordered B2 precipitates, due to their prevalence in high-temperature refractory and multi-principal element alloys \cite{refractories_begley1968high, refractories_osti_4515686, refractories_naka1997designing, precipitates_wang2018coherent}.  However, while some progress has been made in targeted studies \cite{BCC_B2_frey2022high, BCC_B2_frey2024high, BCC_B2_kube2024navigating, BCC_B2_li2020phase, BCC_B2_ma2017bcc, BCC_B2_shaysultanov2017novel, BCC_B2_wang2022phase, BCC_B2_whitfield2020effect}, the enormity of the design space for BCC/B2 alloys strongly motivates the use of artificial intelligence for discovery.

Historically, the development process for new alloys has been slow, often requiring more than a decade, due to complex iterative experimental loops.  Recent advances in \textit{ab-initio} simulations, such as density functional theory (DFT) \cite{kohn1996density} and molecular dynamics \cite{DFT_bartolotti1996introduction, DFT_geerlings2003conceptual, Visual_MD_humphrey1996vmd, VASP_kresse1996software, VASP_kresse1996efficiency, VASP_kresse1993ab}, have accelerated the materials discovery and enabled extensive ground-truth databases of stable compounds (typically containing 3 or more elements)  \cite{AFLOW_curtarolo2012aflow, Jain2013MaterialsProject, OQMD_saal2013materials}. However, the properties of such multi-element alloys depend on beyond-atomistic level dynamics. Computational alloy discovery relies more on thermodynamic simulation methods such as CALPHAD (CALculation of PHAse Diagrams). CALPHAD uses bulk-scale calculations of competing free energy curves to determine the material phases that will be stable at a given temperature and composition. CALPHAD has been applied to alloy development as early as the 1970s \cite{Calphad_kaufman1974calculation}, and modern software packages such as Thermo-Calc \cite{andersson2002thermocalc} make high-throughput calculations for alloy screening relatively straightforward. Simulations like DFT and CALPHAD are commonly used as feedback for algorithmic optimization loops such as Bayesian Optimization \cite{vela_data-augmented_2023,hastings_accelerated_2025}.

\paragraph{Language models for materials}

Most recent AI-driven materials discovery efforts use graph neural networks (GNNs), which excel as discriminative predictors from structured representations (forward design). \citet{merchant2023scaling} exemplifies the forward design approach, employing a greedy algorithm to generate candidate compounds, which are then evaluated for thermodynamic stability using a GNN. Several other studies explore the application of GNNs to predict material properties~\cite{chen2023md, chen2019graph}. %since it reduces the time complexity by many factors compared to classical algorithmic approaches like DFT (density function theory).
By contrast, inverse design begins with a target set of properties and aims to generate novel material candidates expected to exhibit those properties. \citet{gruver2024fine} demonstrates that local LMs can be fine-tuned from a dataset of stable crystals to produce novel stable crystals, similarly to \cite{antunes_crystal_2024}. This focus on thermodynamic stability has characterized most other recent work this area~\cite{sriram2024flowllm, yang2024generative, yan2024invariant}. Perhaps the most similar recent paper to the present effort, PLaID \cite{xu_plaid_2025}, applies DPO to Llama-7b to improve stability of generated crystals. However, as \citet{Seshadri2024GNoMECritique} note, generating thousands of stable materials is not practically useful for working materials scientists. Another limitation of many of previous studies is the use the crystallographic information file (CIF) format, which has both intrinsic downsides~\cite{xiao2023invertible} and little direct representation in e.g., scientific literature, raising questions about what useful biases LMs can bring to CIF-based inverse design tasks. Other recent work leverages LMs without doing any parametric optimization, often via agentic approaches~\cite{lai2025prim, gan2025large, yang2024generative}.

%Some other works also exist that tried a more agentic approach for material design.

%Other work also mentions LM integration in self-driving materials lab as optimizing agent making decisions based on prior outcomes~\cite{abolhasani2023rise}, or a flexible agentic system capable of improving materials science skill of any LM~\cite{zhang2024honeycomb}.

%We could not find any existing research that explores the impact of RLHF-style preference optimization on material composition generation. This is a complex problem because it is a multi-objective optimization problem with the objectives being: 1) generating stable compositions; 2) generating compositions with some desirable properties beyond stability; and 3) generating compositions from a defined search space without compromising the diversity of elements.

\section{Method}

\begin{figure*}
    \centering
    \includegraphics[width=1\linewidth]{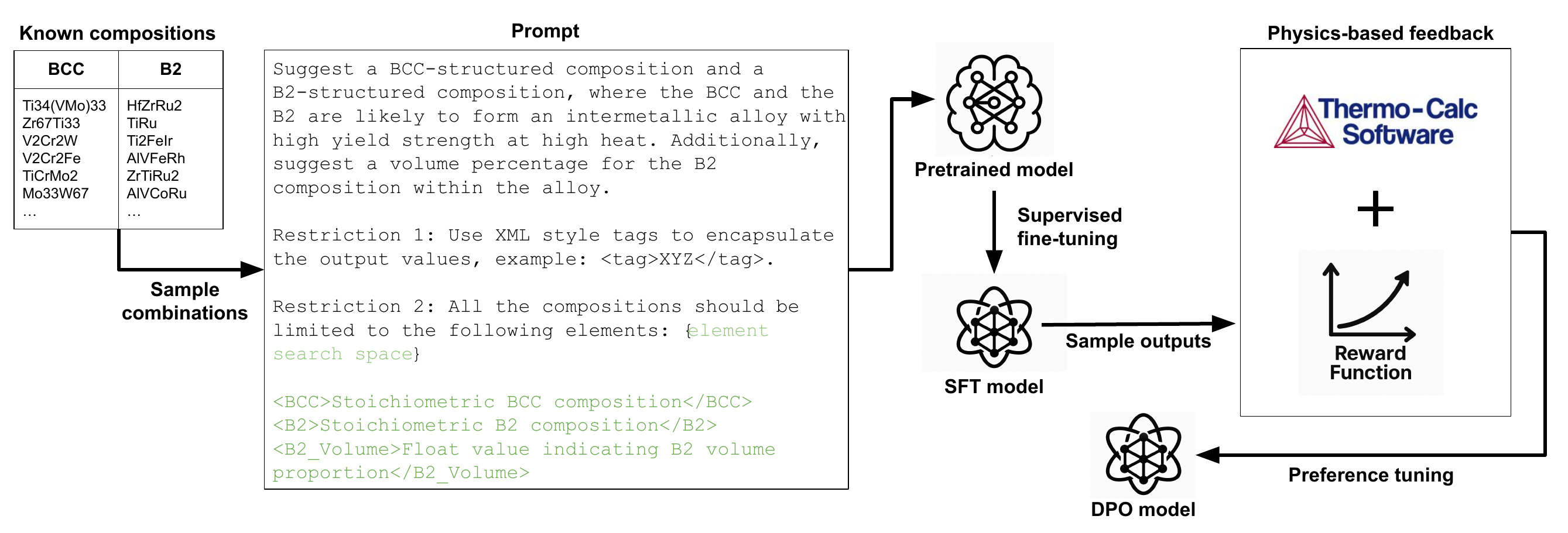}
    \caption{Schematic representation of training the language model for alloy design starting from a pre-trained language model using SFT, physics-based feedback, and DPO.}
    \label{fig:schematic_diagram}
\end{figure*}

% The goal of our work is to optimize parameters of a model for generating \triple triples constituting alloys with potentially strong properties at high temperature. For this we perform two-step parametric optimization using supervised fine-tuning (SFT), followed by direct preference optimization (DPO). The samples from the SFT model go through an automated feedback and reward function before we perform DPO to further refine the quality of generations. The training input to the model is an instruction with a detailed overview of the task followed by two restrictions and the ideal outcome, i.e., a triplet of BCC composition, a B2 intermetallic and the volume percentage of the B2 intermetallic in the alloy as in Figure~\ref{fig:schematic_diagram}.

Prior work has shown that fine-tuned language models can generate CIF files of stable inorganic materials~\cite{gruver2024fine}. However, alloy design is a more nuanced problem that involves satisfying multiple objectives beyond stability. In this work, we focus on a specific class of alloy structures, unlike earlier efforts that search with the sole objective that the crystal is thermodynamically stable. As a result, we bypass the use of CIF files and instead focus solely on compositional generation.

Our approach involves three key steps. First, we construct a cold-start dataset of \triple triples and use it to train a model via supervised fine-tuning (SFT), enabling it to explore the full alloy design space. Next, we sample \triple candidates from the model and evaluate them using Thermo-Calc for thermodynamic feedback. Finally, we use this feedback to define a hierarchical reward function based on synthesize-ability, and apply Direct Preference Optimization (DPO) to align the model with expert-guided preferences. Figure~\ref{fig:schematic_diagram} illustrates this pipeline.

Rather than optimizing each objective independently, we incorporate expert preference into the reward formulation to enable multi-objective optimization through a unified learning signal.

\paragraph{Supervised fine-tuning (SFT)}
% \paragraph{Supervised fine-tuning} 

To build our dataset for SFT, a list of 207 known BCC and 88 known B2 compositions was collected from the Materials Project (CC-BY 4.0)\cite{Jain2013MaterialsProject} and filtered based on stability and alloying suitability \cite{andersson2002thermocalc, tchea7thermocalc}. These elements and compounds were then interchanged in various soluble percentages and verified using the Thermo-Calc (SUNLL) \cite{andersson2002thermocalc} TCHEA7 database (DSUNLL) \cite{tchea7thermocalc} to produce ground-truth triplets of the form \triple. The SFT dataset consists of all possible combinations of these compositions (18,216 distinct pairs), combined with three B2 volume percentages for each, sampled from a normal distribution with a mean of .45, capped at .20 and .70, for a total size of 54,648 examples.
% This dataset was randomly split into 80 and 20 percent fractions for training and validation, respectively. 
Additional details can be found in Appendix \ref{appendix_sft}. 

We tune the SFT model using a causal language modeling (CLM) objective, using an instruction-based prompt (Figure \ref{fig:schematic_diagram}). To reduce the number of trainable parameters, we employ low-rank adapter modules (LoRA)~\cite{hu2022lora}, configuring the adapters with a rank of 8 and scaling factor $\alpha$ = 32. This setup results in only ~0.027\% (for LLaMA) and ~0.057\% (for OLMo) of parameters being updated during fine-tuning. Following ~\citet{gruver2024fine}, we introduced special tokens to the tokenizer vocabulary (if they did not exist) for padding, beginning of sentence, end of sentence, and unknown to properly tokenize chemical formulas. More details about the training can be found in Appendix~\ref{appendix_sft_data}. The generations sampled from this stage is combined into a master composition based on the molar volume percentage of B2 and fed to Thermo-Calc.

\paragraph{Reward function}
\label{sec:reward_function}
% \sam{What is a BCC/B2 alloy, what is thermocalc}
% \sam{Diagram or figure illustrating the raw output from thermocalc}
% \neal{Putting in some text to answer your questions, feel free to nuke it}

% For an alloy to be useful, it must have acceptable properties to meet a performance goal and be easily fabricable. 
Preference feedback for DPO comes from the Thermo-Calc tool~\cite{andersson2002thermocalc}, which takes as input a single composition and temperature and, using a combination of simulation and databases of empirical results, predicts what phases are likely to exist in what quantity at that temperature. To create a reward score for an SFT-generated \triple triple, we use the B2 volume \% to combine the BCC and B2 compositions into a single master composition, then query Thermo-Calc on this composition at a range of temperatures from $373 K$ to $2273 K$. An example of output from Thermo-Calc is shown in Figure~\ref{fig:thermocalc_op}.

% The advantageous properties of BCC/B2 superalloys arise from the combination of strong but brittle B2 precipitates embedded within a deformable BCC matrix. 
Realizing a fabricable superalloy requires multiple interplaying factors to align during processing, namely: (i) the BCC phase must be the first to solidify from a liquid melt; (ii) the B2 phase should form at a temperature below that of the BCC phase, but still at as high of a temperature as possible to maximize the thermal operation limit of the alloy; (iii) the alloy must be comprised entirely (or nearly entirely) of BCC and B2, as other intermetallic compounds are often brittle and weak, making them largely undesirable; and (iv) the BCC and B2 phases should have nearly identical crystal lattice sizes, which reduces the build-up of internal stresses in the alloy during processing and use. We operationalize these viability rules as follows (in descending order of importance):
\begin{enumerate}[noitemsep,nolistsep,topsep=0pt,leftmargin=20pt]
    \item There must be some temperature at which both a solid BCC and B2 phase exist simultaneously. (\texttt{bcc\_b2\_exist})
    \item The BCC must form first as the temperature decreases. (\texttt{bcc\_forms\_first})
    \item A B2 phase must exist close to room temperature, $373 K$. (\texttt{b2\_room\_temp})
    \item No more than 10\% of non BCC/B2 phases should form at any temperature. (\texttt{others\_exceed\_10\%})
\end{enumerate}

% \begin{enumerate}[noitemsep,nolistsep,topsep=0pt,leftmargin=20pt]
%     \item There must be some temperature at which both a solid BCC and a B2 exist simultaneously. ($Objective_1$)
%     \item The BCC must form first as the temperature decreases, as BCC/B2 superalloys form from ``pockets'' of liquid B2 solidifying inside an existing BCC matrix. ($Objective_2$)
%     \item A B2 phase must exist at room temperature, $373^{\circ}K$. ($Objective_3$)
%     \item No more than 10\% of non BCC/B2 phases should form at any point. ($Objective_4$)
% \end{enumerate}

When all these criteria are satisfied, the quality of a candidate is measured as the minimum difference in lattice parameter (reported in Å) between BCC and B2 phases at any temperature (\texttt{min\_lattice\_mismatch}). This mismatch value typically varies from $10^{-1}$ to $10^{-7}$. The overall reward is numericized as a weighted sum of indicators for these boolean conditions:
% $$
\begin{align}
\label{eq:reward}
& \text{Reward}(\text{BCC}, \text{B2}, \text{Volume})  = \notag\\  
&   -1000\, \mathbf{1}_{\neg\texttt{bcc\_b2\_exist}}
    - 100\, \mathbf{1}_{\neg\texttt{bcc\_forms\_first}} \notag \\
&   - 10\, \mathbf{1}_{\neg\texttt{b2\_room\_temp}} 
    - \mathbf{1}_\texttt{others\_exceed\_10\%} \notag \\
&   - \texttt{min\_lattice\_mismatch}
\end{align}
% $$

% where, $Objective_1$ is bcc\textunderscore b2\textunderscore exist, $Objective_2$ is bcc\textunderscore forms\textunderscore first, $Objective_3$ is b2\textunderscore room\textunderscore temp, \& $Objective_4$ is others\textunderscore exceed\textunderscore 10\%.

% \begin{align}
%     R( \textit{Thermo-Calc}(\text{BCC}, \text{B2}, \text{Volume})) &= -k_1 \cdot \text{Objective}_1 
%     - k_2 \cdot \text{Objective}_2 \notag \\
%     &\quad - k_3 \cdot \text{Objective}_3 
%     - k_4 \cdot \text{Objective}_4 \notag \\
%     &\quad - \text{minimum lattice mismatch}
% \end{align}
% \begin{center}
% \text{where $k_1$ = 1000, $k_2$ = 100, $k_3$ = 10, and $k_4$ = 1}
% \end{center}

The reward (or regret) score ends up negative log-scaled, with a worst possible score of $\sim-10^{3}$ and best of $\sim-10^{-7}$, with $>-10^{0}$ being the viability threshold of obeying the four basic rules. These coefficients reflect a tiered prioritization of synthesis realism: thermodynamic coexistence is fundamental, while lattice mismatch offers fine-grained selection. Ultimately the score reflects the \textbf{viability and potential for favorable properties of the candidate BCC/B2 alloy}, rather than a direct estimate of its properties per se. This reflects the way CALPHAD calculations are used in traditional alloy design (e.g. \cite{holgate_design_2025}), as a screening filter on potential candidates. 
Since this class of materials is in its infancy, the reward function does not target high-temperature performance, instead focusing on candidates with favorable properties at any temperature in range. It could easily be made more specific by, for instance, setting a minimum temperature threshold on the various rules to ensure that they hold at the target conditions.

\paragraph{Direct preference optimization (DPO)}

To guide our model toward producing higher-quality \triple triples, we sample candidates $S_{\theta_{SFT}}$ from the SFT model and calculate their reward score using Eq. \ref{eq:reward}. From the output of our reward function we create a pairwise preference dataset $\mathcal{D_\text{DPO}}(y^+, y^-)$, where $y \in S_{\theta_{SFT}}$ indicating a preferred generation $(y^+)$ over $(y^-)$. We want to push our model towards a region of higher rewards by optimizing a contrastive objective, reviewed more fully in the appendix, where hyperparameter $\beta$ controls the distance between the distribution of the original SFT model distribution and that of the the new model. 
We want the internal reward mapping of the model (as no separate reward model is required in DPO) to learn from our multiobjective reward scores and push the model to search the parametric space of higher average reward. However, to prevent the preference tuned model from going wildly out of distribution or hacking the reward function~\citep{rafailov2024scaling}, we set $\beta = 0.5$. Training was conducted using a low-rank adapter module, trained for 1 epoch (more details in \ref{appendix_dpo}).

For the DPO dataset, we sample 5,000 \triple triples from the SFT model, then use Thermo-Calc to compute a scalar reward for each generation. We construct a preference dataset with the top 25\% generations, as ranked by reward, paired with 100 randomly selected lower ranked generations. This strategy allows the model to learn from relative preferences, encouraging discrimination between high- and low-quality outputs.

%We constructed a preference dataset: the top 25\% generations, as ranked by reward, were each paired with 100 randomly selected lower ranked generations. This strategy allows the model to learn from relative preferences, encouraging discrimination between high- and low-quality outputs (statistical rejection TKTK).

%Using this preference dataset, we applied Direct Preference Optimization (DPO TKTK), enabling the model to optimize the parameters using contrastive learning and pushing the model generations towards higher reward region. We observed a marked increase in the scalar reward post-training (see Fig. TKTK), indicating improved generation quality.

%Training was conducted with a single-turn reinforcement learning  using a low-rank adapter module, trained for one epoch. Further details of the training protocol, hyperparameters, and implementation considerations are provided in the Appendix.

\section{Experiment}

% \paragraph{Baselines} We use two baselines for comparison with our SFT and DPO models:
% \begin{itemize}[noitemsep,nolistsep,topsep=0pt,leftmargin=10pt]
%     \item \textbf{Random search:} Conventional alloy discovery approaches often do parametric sweeps of composition space for promising candidates. We approximate this approach by constructing a grid of BCC- and B2-forming elements and sample random compositions from it. The sampling of both compositional elements and B2 volume fractions is done randomly, and the list of elements included both stable and metastable BCC- and B2-formers.
%     % The easiest and common would be to just randomly sample BCC-B2 pairs from the defined element space and sample B2 volume from a normal distribution between 20-70\%.
%     \item \textbf{API-based LMs:} 
%     % Some of the state-of-the-art language models are opaque and can be accessed only through API's via prompting. 
%     Our second baseline consists of one-shot and few-shot prompting of three state-of-the-art proprietary API-based models: Gemini-2.5, GPT-4.1 and GPT-o3. We find zero-shot prompting from these models unreliable in terms of output format, and do not include this as a condition. In the one-shot setting we randomly sample a single exemplar from the SFT model output. In the few-shot setting we provide top 10 and bottom 10 generations from the SFT model as exemplars, ranked on reward. Exact prompts and exemplars are provided in Appendix~\ref{API_prompt}.
%     %This is provided to the model with context of `good' or `bad' generation examples. 
    
% \end{itemize}

\paragraph{SFT and DPO models} We perform SFT and DPO on three open instruction-tuned LMs of comparable size: LLaMA-3.1-8B~\cite{grattafiori2024LLaMA}, Gemma-2 (9B)~\cite{team2024gemma}, and OLMo-2-7B~\cite{olmo20242}. We use low-rank adapters ($\alpha = 32$, $rank = 8$) for training, with 8-bit quantized models.

% The primary differences between these two models are: (1) pre-training data, \& (2) vocabulary size. Other than that, these models have a similar training regime, with OLMo using PPO to perform Reinforcement Learning from Verifiable Rewards (RLVR) during the post-training stage on top of their DPO trained model.

\paragraph{Baselines} 
%We use two types of baselines for comparing our SFT and DPO models. The first is a \textit{random search}, which is an approximation of a parametric sweep. The second approach is to use \textit{API based models} (GPT 4.1, GPT o3, and Gemini 2.5) with one-shot and few-shot prompting using exemplars sampled from the DPO training data. We additionally try a ``guided'' variant of GPT-4.1 with the 207 BCCs and 88 B2s identified as potential options. More details of the baselines can be found in Appendix~\ref{baseline}.

To properly evaluate the gains and limitations of our approach, we compare it against several varyingly strong baselines. \textbf{(1) Random search}: Alloy design has traditionally been a serendipitous process; accordingly, one of our baselines involves randomly searching the BCC/B2 composition space, with the B2 molar volume sampled uniformly between 20\% and 70\% (more details in Appendix~\ref{baseline}). 
\textbf{(2) Prompting API-based models}: We use few-shot prompting of state-of-the-art (at the time of writing) API-based large LMs, including GPT-4.1, GPT-O3, and Gemini-2.5. Prompts are available in the Appendix. \textbf{(3) Prompt tuning}: We find empirically (see below) that prompting approaches suffer from poor diversity in their outputs. To create a stronger baseline, we extend the most balanced API model (Gemini-2.5) and automatically tune the input prompt to encourage diversity, using the MIPROv2 optimization method from the DSPy library \cite{khattab2023dspy}. \textbf{(4) Agentic approach}: To further investigate the capabilities of the API based models we create a simple agentic system where two agents: a generator and an evaluator work in conjunction to come up with high quality alloy compositions. The generator agent generates a composition and the evaluator accepts or rejects the composition with a feedback. We optimize the generator via verbal reinforcement from the evaluator agent. More details in Appendix. \textbf{(5) Prior published models}: Additionally, we incorporate generations from previously published generative models, including Crystal-LLM~\cite{gruver2024fine} and CDVAE~\cite{xie2021crystal}, which aim to generate crystal structures of inorganic compounds. Although these models are trained for general-purpose stable inorganic crystals, we filter their outputs to retain only those compositions that fall within our target alloy design space, i.e., potential BCC/B2 alloy composed of TCHEA elements.

\begin{table*}[ht]
    \small
	\centering
\begin{tabular}{lcccccl}
\hline
\textbf{Model} &
  \textbf{Validity} &
  \textbf{\begin{tabular}[c]{@{}c@{}}Coverage \\ Recall\end{tabular}} &
  \textbf{\begin{tabular}[c]{@{}c@{}}Coverage \\ Precision\end{tabular}} &
  \textbf{Novelty} &
  \textbf{\begin{tabular}[c]{@{}c@{}}Mean \\ Reward\end{tabular}} &
  \textbf{\begin{tabular}[c]{@{}l@{}}Unique\\ pairs @100\end{tabular}} \\ \hline
Random search           & 0.80 & 0.98 & 0.82 & 0.44 & -883.71 & 1.0                      \\ \hline
CDVAE                   & 0.73 & 0.43 & 0.07 & 0.94 & --      & --                       \\
Crystal-LLM-7B          & 0.90 & 0.34 & 0.18 & 0.80 & --      & --                       \\
Crystal-LLM-13B         & 0.87 & 0.44 & 0.17 & 0.81 & --      & --                       \\
Crystal-LLM-70B         & 0.91 & 0.45 & 0.17 & 0.83 & --      & --                       \\ \hline
GPT-4.1                 & 1.00 & 0.32 & 1.00 & 0.86 & -53.23  & 0.44                     \\
GPT-O3                  & 1.00 & 0.42 & 1.00 & 0.99 & -75.43  & 0.66                     \\
Gemini-2.5              & 0.99 & 0.79 & 0.99 & 0.81 & -106.22 & 0.82                     \\
Prompt-tuned Gemini-2.5 & 0.99 & 0.83 & 1.00 & 0.98 & -350.34 & 0.91                     \\ \hline
Agentic GPT-4.1         & 1.00 & 0.48 & 1.00 & 0.77 & -542.60    & 0.98 \\
Agentic Gemini-2.5      & 1.00 & 0.78 & 1.00 & 0.87 & -19.87    & 0.61 \\ \hline
Gemma SFT               & 0.99 & 0.99 & 1.00 & 0.94 & -220.41 & 0.98                     \\
Llama SFT               & 0.99 & 0.99 & 0.99 & 0.92 & -215.92 & 0.99                     \\
OLMo SFT                & 0.99 & 0.99 & 0.99 & 0.92 & -218.54 & 1.00                     \\ \hline
Gemma DPO               & 1.00 & 0.95 & 1.00 & 0.97 & -206.71 & 0.92                     \\
Llama DPO               & 0.99 & 0.98 & 1.00 & 0.93 & -175.89 & 1.00                     \\
OLMo DPO                & 0.99 & 0.98 & 1.00 & 0.95 & -268.72 & 0.98                     \\ \hline
\end{tabular}
	\caption{Evaluation of generative models on validity, coverage, and novelty as proposed by \citet{xie2021crystal}, as well as mean reward score and what fraction of 100 generated BCC/B2 pairs are unique (lower indicates more self-repetition).}
	\label{tab:basic_results}
\end{table*}
\vspace{-\baselineskip}

\section{Evaluation}
\subsection{Basic Results}
\label{sec:basic_results}
Our basic results, shown in Table \ref{tab:basic_results}, use compositional validity, coverage, and novelty metrics, as introduced by \citet{xie2021crystal} and later adopted by \citet{gruver2024fine}. Compositional validity is assessed using the Pauling electronegativity test, which ensures that the constituent elements exhibit appropriate electronegativity differences~\cite{davies2016computational}. Coverage is computed as the Euclidean distance between the normalized feature vectors of generated compositions and all 18,216 potential BCC/B2 alloy compositions--coverage recall measuring what percentage of the space is produced, and coverage precision measuring what percentage of produced compositions belong within the space. Novelty is measured as the pairwise distance between generated samples and all known (existing) alloys containing two or more TCHEA elements, based on their feature representations. While coverage measures how well the generated compositions span the known design space, novelty captures how different they are from all existing alloys. We also report mean reward score among generated compositions, and ``Unique pairs @100'', the fraction of 100 generated BCC/B2 pairs that are unique. A lower score on this latter value indicates more self-repetition and less diversity. Following prior work, we use Matminer~\cite{ward2018matminer} to vectorize the compositions. We sample at least 1000 generations from each model with $\tau=1.0$. 
\textbf{An ideal model should have near-perfect validity and achieve a balance between coverage, novelty and reward.}

 From Table \ref{tab:basic_results}, we observe that general-purpose crystal generation models struggle to produce valid BCC/B2 alloys within our defined design space. These models show low coverage recall and precision, frequently missing key regions of the space and generating chemically irrelevant compositions, over half of which fail the compositional validity checks. Randomly sampling from existing BCC and B2 compositions leads to a high coverage but the final result is often (about 30\% times) not a valid composition and not a BCC/B2 alloy for about 20\% of the time. Novelty also goes down since they are similar to existing alloys in the MP database.
 
 Among the API-based models, the generated compositions demonstrate high validity and coverage precision, often near perfect. However, they exhibit low coverage recall and low pair uniqueness, meaning that they tend to repeat themselves while failing to fully span the design space. Their relatively high novelty scores indicate they produce compositions distinct from those in the Materials Project database. They produce high-reward candidates, especially GPT-4.1, indicating that their retrieved/parametric knowledge provides useful biases, though these biases presumably also prevent them from exploring certain regions of the design space, hence the lower coverage. The prompt-tuned Gemini-2.5 model, whose prompt is optimized toward generating diverse outputs, demonstrates higher coverage and pair uniqueness than the other API-based models, but this comes at the cost of reward, with its proposed alloys underperforming even the SFT models, which are not tuned for reward. 

The local SFT models, trained on a uniform sample of \triple triples, are all comparable. They demonstrate high validity, coverage, novelty and pair uniqueness. This indicates that they succeed at becoming a ``blank slate'', generating uniformly from the designated space of possible \triple triples. While this doesn't make them very useful alloy-proposers on their own, it does make them suitable for further optimization toward a specific goal, which we implement in the form of DPO. 

% By contrast, the local models demonstrate uniformly good coverage of the design space. Notably, the preference-tuned (PT) models exhibit higher novelty than their supervised fine-tuning (SFT) counterparts, while simultaneously showing lower coverage recall. This suggests that PT models are more hyper-focused on specific regions of the design space compared to the broader exploration of SFT models. However, this analysis alone does not fully capture the distinction between PT and SFT models. To further differentiate them, we conduct a second stage of evaluation, where we assess the synthesizability of the generated alloys using thermodynamic calculations.

\subsection{Effect of preference tuning}
% We conduct this analysis to determine whether preference tuning is necessary for niche design tasks such as this. If the generations from the PT model are less likely to be synthesizable (based on thermodynamic calculations of phases) in a materials laboratory, it would suggest that the reward signal was not meaningful to the model. In that case, the computational overhead of DPO could be avoided in the future.

Table \ref{tab:basic_results} shows that the DPO models, with the exception of OLMo, show a modest improvement in mean reward over their SFT precursors, while maintaining their high coverage of the design space and generated pair uniqueness. Their mean reward is lower than that of the API based models (excluding prompt-tuned Gemini-2.5), indicating that they learn fewer biases than these larger models.

\begin{figure}[htbp]
  \centering

  % Left figure
  \begin{minipage}[t]{0.48\textwidth}
    \centering
    \includegraphics[width=\linewidth]{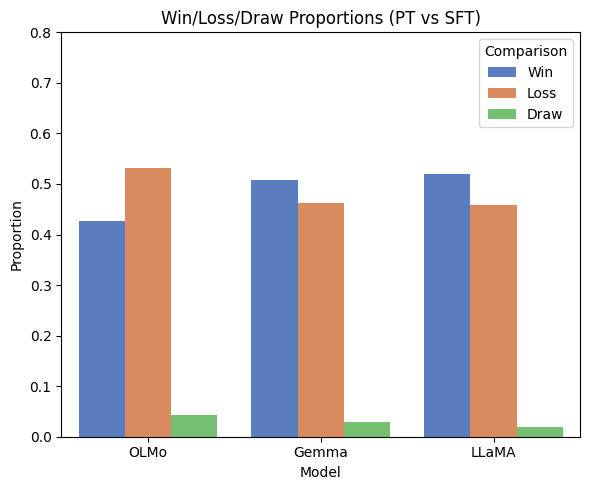}
    \caption{Each bar represents the proportion of cases where the DPO model outperformed (Win), underperformed (Loss), or matched (Draw) its SFT counterpart in reward score. 
    % While DPO models generally achieve more wins than losses for Gemma and LLaMA, OLMo exhibits a higher loss rate, indicating that preference tuning may not consistently benefit all model architectures.
    }
    \label{fig:win_draw}
  \end{minipage}
  \hfill
  % Right figure
  \begin{minipage}[t]{0.48\textwidth}
    \centering
	\includegraphics[width=\linewidth]{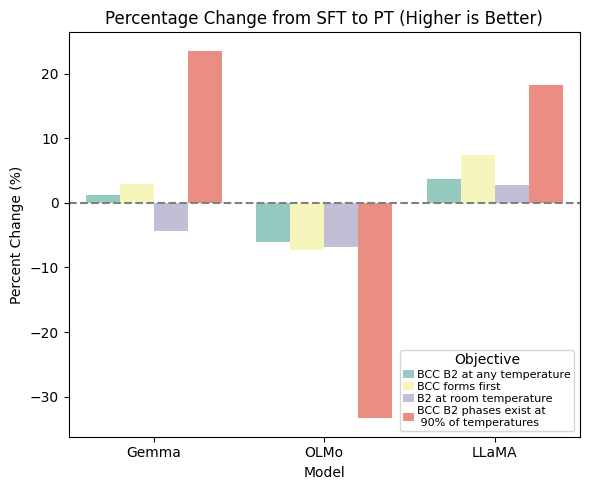}
	\caption{Percentage change in objective satisfaction from SFT to DPO models across Gemma, OLMo, and LLaMA. The plot illustrates the relative improvement or degradation in meeting four alloy design objectives after preference tuning (DPO). 
    % While Gemma and LLaMA show consistent improvements, particularly in achieving phase stability across 90\% of temperatures, OLMo experiences a notable decline across all objectives, suggesting that preference tuning may have pushed its generations off-distribution.
    }
	\label{fig:objectivecomparison}
  \end{minipage}

\end{figure}

% The first step is determining whether our optimization worked and the PT model generations outperform their SFT counterpart more often than they underperform. To discretely understand how they compare in a head-to-head comparison, we
Figure~\ref{fig:win_draw} illustrates the effect of DPO with Win/Draw/Loss analysis based on reward score. Gemma and LLaMA DPO models win 49.8\% and 52.1\% of the time and lose 46.1\% and 45.4\% of the time, respectively. The rest were draws. However, the OLMo DPO model lost to its SFT counterpart 52.4\% of the time and won only 42.3\% of the time.

Figure~\ref{fig:objectivecomparison} assesses how effectively the cumulative learning signal optimized the models for individual synthesis objectives. We evaluate the four manually-chosen subcomponents of the reward function: (1) BCC and B2 phases must coexist at some temperature; (2) BCC must form first at a higher temperature; (3) B2 must exist at room temperature; and (4) BCC/B2 phases must be present across 90\% of the evaluated temperature range. We compute the percentage change in the satisfaction rate---defined as the proportion of generated alloys that satisfy each objective---from the SFT to the DPO models. As shown, all synthesis objectives improve in LLaMA, while three out of four improve in Gemma. In contrast, OLMo exhibits degradation across all four objectives following preference tuning. Two key insights emerge from these results: (1) optimizing for the presence of the B2 phase at room temperature remains challenging, as both Gemma and OLMo perform worse on this criterion, and LLaMA shows only modest improvement; and (2) combining multiple reward signals in this setup can push certain architectures like OLMo off-distribution, leading to a collapse in performance across objectives, possibly due to its smaller capacity or mismatch with reward distribution. However, the fact that two out of the three models improved after preference tuning, using a reward function derived from practical design objectives, suggests that a similar learning framework with hierarchical reward signals could be an effective way to optimize models.

\subsection{Hyperfixation in API-based models}

% Elemental diversity is a critical aspect of alloy design, as the inclusion of a wide range of chemical elements enables fine-tuning of properties, enhances phase stability, and suppresses the formation of undesirable phases. Consequently, experts prefer models that can explore a broad elemental space.

API-based models such as GPT-4.1 and Gemini-2.5 models are powerful and easy to use, which begs the question of whether local models have a place in LM-driven materials discovery alongside API-based models and the agentic systems built on top of them. 
% quite convenient due to their easy interactive platforms and negligible computational overhead from the users end. So, experts may consider using API-based models (like Gemini-2.5) as a viable alternative since it does not require much labor to generate alloy compositions. While API-based models demonstrated reasonable performance, 
Our analysis in Section \ref{sec:basic_results} shows that the strong biases of these models limits their coverage recall and generated BCC/B2 pair uniqueness. To better understand this limitation, we conduct a focused analysis to understand patterns of hyperfixation in their behavior.

\begin{table*}[h]
\centering
\scriptsize
\setlength{\tabcolsep}{3pt} % tighter column spacing
\resizebox{\textwidth}{!}{%
\begin{tabular}{@{}lcrcrcrcrcrcr@{}}
\toprule
\multirow{2}{*}{\textbf{Rank}} & \multicolumn{2}{c}{\textbf{GPT-4.1}} & \multicolumn{2}{c}{\textbf{GPT-O3}} & \multicolumn{2}{c}{\textbf{Gemini-2.5}} & \multicolumn{2}{c}{\textbf{Prompt-tuned Gemini-2.5}} & \multicolumn{2}{c}{\textbf{Llama DPO}} & \multicolumn{2}{c}{\textbf{Llama SFT}} \\
\cmidrule(l){2-13}
& \textbf{Elements} & \textbf{Freq} & \textbf{Elements} & \textbf{Freq} & \textbf{Elements} & \textbf{Freq} & \textbf{Elements} & \textbf{Freq} & \textbf{Elements} & \textbf{Freq} & \textbf{Elements} & \textbf{Freq} \\
\midrule
\textbf{1} & \{Mo, Nb\} & 0.500 & \{Mo, Nb, W\} & 0.578 & \{Mo, Nb\} & 0.145 & \{Mo, Nb, Ta\} & 0.115 & \{Mo, Nb, Ti\} & 0.072 & \{Cr, Ti, V\} & 0.041 \\
\textbf{2} & \{Nb, W\} & 0.382 & \{Mo, Nb, Ta, W\} & 0.152 & \{Mo, Nb, W\} & 0.136 & \{Mo, Nb, Ti\} & 0.096 & \{Mo, Nb, W\} & 0.048 & \{Ti, V, W\} & 0.038 \\
\textbf{3} & \{Mo, Nb, W\} & 0.105 & \{Mo, Ta, W\} & 0.140 & \{Nb, W\} & 0.089 & \{Mo, Nb, Ta, Ti\} & 0.059 & \{Nb, Ti, W\} & 0.048 & \{Nb, Ti, V\} & 0.037 \\
\textbf{4} & \{Cr, Mo, W\} & 0.008 & \{Mo, Nb, V, W\} & 0.045 & \{Nb, Ta, W\} & 0.073 & \{Mo, Nb, Ta, W\} & 0.054 & \{Mo, Ti, W\} & 0.046 & \{Mo, Ti, V\} & 0.036 \\
\textbf{5} & \{Mo, Nb, Ta\} & 0.001 & \{Mo, Nb, Ta\} & 0.020 & \{Cr, Mo, W\} & 0.062 & \{Mo, Ta, W\} & 0.052 & \{Cr, Mo\} & 0.040 & \{Mo, Nb, W\} & 0.033 \\
\bottomrule
\end{tabular}%
}
\caption{Top 5 most frequent BCC element combinations generated by each model.}
\label{tab:topk_table_new}
\end{table*}

Table \ref{tab:topk_table_new} explains the prompting model result by showing the top 5 BCC element combinations generated by a selection of models. We can see that half of few-shot GPT-4.1's BCCs are Mo/Nb combinations, and 98\% use some subset of Mo/Nb/W. Few-shot Gemini shows a similar but less extreme level of fixation, with at least 36\% of its BCC candidates a subset of the same Mo/Nb/W combination. A prompt-tuned Gemini-2.5 few-shot approach reduced this even more, with about 13\% BCC with some combination of Mo/Nb/Ta. By contrast, DPO LLaMA shows a much more even spread, only slightly more concentrated than SFT LLaMA. This means that the API models achieve high average reward by fixating on a small selection of elements and element combinations.

\begin{figure*}[h]
	%\vspace{-10pt}
	\centering
	\includegraphics[width=.8\textwidth]{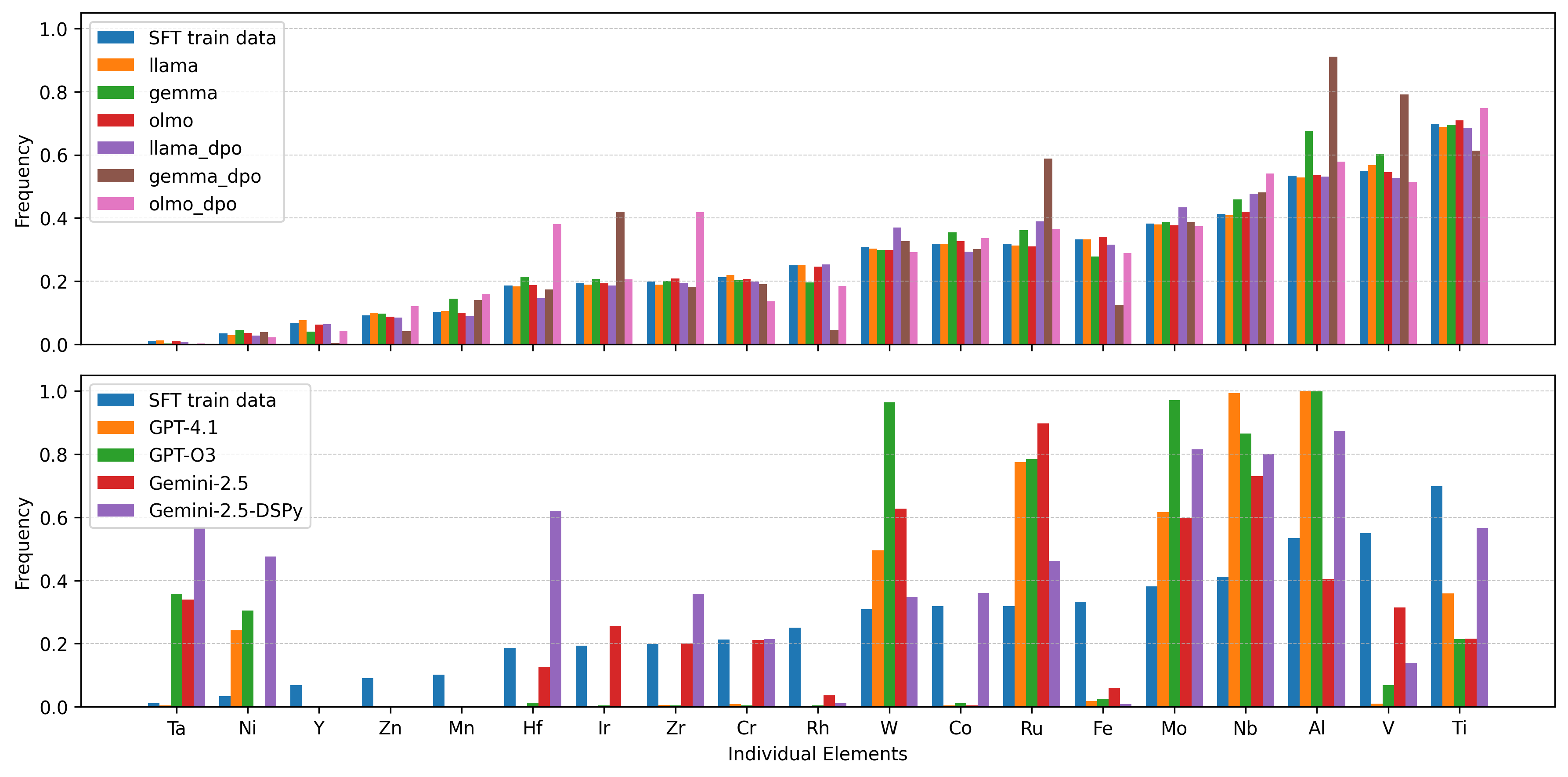}
	\caption{Output frequencies of individual elements by trained models (top) and API models (bottom), respectively, compared to the training data.} 
	\label{fig:dpo_fs_element_frequencies}
	% \vspace{-8pt}
\end{figure*}

Finally, Figure \ref{fig:dpo_fs_element_frequencies} shows the distribution of individual elements favored by the SFT and DPO models versus the API models. The top plot shows that SFT and DPO generations have an element distribution similar to the training data. Among all the trained models we can see that DPO Gemma and DPO OLMo are fixating slightly more on some elements like Ir/Ru/Al/V and Hf/Zr/Nb/Ti, respectively. In particular DPO OLMo generated Hf and Zr at much higher frequency and Ir by DPO Gemma than the training compositions. The bottom plot shows the fixation of few-shot GPT-4.1 (green), Gemini-2.5 (red) and prompt-tuned Gemini-2.5 (violet) on certain elements like Ta/Ni/Hf/Zr/W while completely missing on elements like Y/Zn/Mn. Gemini is noticeably more adherent to the training data element frequencies than GPT-4.1, with GPT-4.1 hyperfixating on Nb and Al beyond what is in the training data. 

The sum total of these results shows that API-based models achieve high reward by focusing on known high-reward regions, to the exclusion of unknown regions, and that this behavior is difficult to dislodge via prompt tuning without badly affecting reward. It is widely acknowledged that pre-existing biases affect and limit exploratory materials development \cite{jia_anthropogenic_2019, horgan_biased_2021}, and our analysis seems to indicate that API-based models reflect those same biases. Therefore, there may be a role for models capable of learning useful reward signals while still retaining a high degree of exploratory openness, as our DPO-tuned models demonstrate. 

\section{Discussion}

% \paragraph{Summary of results}
Preference tuning is valued for its ability to optimize language models toward objectives that are (1) noisy and (2) hard to describe or articulate (such as politeness or humor). That makes it appropriate for optimizing LM-generated materials toward arbitrary physical objectives.

Our results show that DPO is able to produce a modest improvement in average reward while maintaining high diversity in output, for two of three local LMs. OLMo, on the other hand, performed worse after DPO across all objectives. We observe increased divergence of key token logits between SFT and DPO for OLMo, which explains the collapse (more analysis in Appendix~\ref{sec: olmo_fail}). 
% This aligns with the well-known sensitivity of preference tuning in smaller architectures~\citep{rafailov2024scaling}. In this case, preference tuning may have pushed the model off-distribution. 
% This raises an important point: reward design alone is not sufficient; model architecture and robustness play a role in how well preference learning works.
% Our training protocol uses SFT to produce a baseline distribution over a specified design space, in our case \triple triples sampled from a discrete set of known BCCs and B2s. Then it applies DPO from physics-based feedback to orient the model toward higher-reward regions without blinding it completely to lower-average reward regions which might still yield good candidates.
While we apply our SFT-DPO training process to a highly specific design space and reward function, it is 
% This is 
a highly general protocol, and could be applied to any engineering problem capable of using an SFT training set to represent a design space and with a computationally-efficient verifier available over generated candidates. One possible example is battery design, where  open-source tools like PyBaMM~\citep{sulzer2021python} could be used to assess generated candidates.

While model training can identify good regions of feature space, black box optimization (BBO) is more suited to identifying standout candidates within that space. BBO methods such as Bayesian Optimization are a major part of computational alloy discovery \citep{hastings_accelerated_2025, wang_bayesian_2022}, and recent work has sought to combine LMs with Bayesian Optimization as both generators of candidate points and discriminators over generated candidates \citep{liu_large_2024,chang_textttllinbo_2025}. While the useful biases of API-based models makes them more likely to suggest high-reward candidates (when used as generators) and more likely to correctly assess provided candidates (when used as discriminators), their tendency to fixate on certain regions of feature space limits their ability to perform the ``explore'' part of the exploration/exploitation tradeoff in discrete optimization. Tuned local models offer a potential solution to this problem by offering more control over their degree of bias, particularly via the $\beta$ parameter of the DPO process.

\paragraph{Limitations}

One limitation of this work is that the predictions produced by Thermo-Calc and similar tools are not perfect, and become less reliable for many-element compositions in regions for which the tool's databases have poor coverage. Engineering a confidence estimate for external feedback, combined with LM reasoning over external context like prior scientific findings, could be a way of mitigating this issue, as could, in a fully realized modeling pipeline, the inclusion of physical experimentation to verify the predicted properties of key candidates. 
 A higher-level limitation is the question of whether, for downstream DO tasks, a higher-reward baseline distribution is actually needed and worth the investment in time and effort to create. If our ultimate goal is to find a small number of exceptional alloy candidates, it might be more efficient to simply perform a search through the output space of the SFT model. Future work will explore this question.

 \paragraph{Conclusion} We apply preference tuning for the first time to LM-driven inverse design of materials toward functional properties, and propose preference-tuned ``high-reward'' models as an intermediate step toward LM-driven materials discovery. Our supervised fine-tuning is successful, while our preference tuning results are positive, though inconsistent between models. While we apply these ideas specifically to BCC/B2 superalloy discovery, the template we introduce here is general, and could be adapted to any design problem where it is possible to collect medium-scale feedback on model-suggested compositions, such as battery or photovoltaic materials Finally, this work is complimentary with other approaches for LM-guided materials discovery, such as agentic approaches, and could be extended to work as an improved baseline distribution for such methods.

\bibliography{alloy_generation_ref, sam_additional_references, material_refs}

\begin{thebibliography}{63}
\providecommand{\natexlab}[1]{#1}
\providecommand{\url}[1]{\texttt{#1}}
\expandafter\ifx\csname urlstyle\endcsname\relax
  \providecommand{\doi}[1]{doi: #1}\else
  \providecommand{\doi}{doi: \begingroup \urlstyle{rm}\Url}\fi

\bibitem[Andersson et~al.(2002)Andersson, Helander, H{\"o}glund, Shi, and Sundman]{andersson2002thermocalc}
Jan-Olof Andersson, Thomas Helander, Lars H{\"o}glund, Pingfang Shi, and Bo~Sundman.
\newblock Thermo-calc \& dictra, computational tools for materials science.
\newblock \emph{Calphad}, 26\penalty0 (2):\penalty0 273--312, 2002.

\bibitem[Antunes et~al.(2024)Antunes, Butler, and Grau-Crespo]{antunes_crystal_2024}
Luis~M. Antunes, Keith~T. Butler, and Ricardo Grau-Crespo.
\newblock Crystal structure generation with autoregressive large language modeling.
\newblock \emph{Nature Communications}, 15\penalty0 (1):\penalty0 10570, December 2024.
\newblock ISSN 2041-1723.
\newblock \doi{10.1038/s41467-024-54639-7}.
\newblock URL \url{https://www.nature.com/articles/s41467-024-54639-7}.
\newblock Publisher: Nature Publishing Group.

\bibitem[Bartolotti and Flurchick(1996)]{DFT_bartolotti1996introduction}
Libero~J Bartolotti and Ken Flurchick.
\newblock An introduction to density functional theory.
\newblock \emph{Reviews in computational chemistry}, pages 187--216, 1996.

\bibitem[Begley et~al.(1968)Begley, Harrod, and Gold]{refractories_begley1968high}
RT~Begley, DL~Harrod, and RE~Gold.
\newblock High temperature creep and fracture behavior of the refractory metals.
\newblock In \emph{Refractory Metal Alloys Metallurgy and Technology: Proceedings of a Symposium on Metallurgy and Technology of Refractory Metals held in Washington, DC, April 25--26, 1968. Sponsored by the Refractory Metals Committee, Institute of Metals Division, The Metallurgical Society of AIME and the National Aeronautics and Space Administration, Washington, DC}, pages 41--83. Springer, 1968.

\bibitem[Brodnik et~al.(2023)Brodnik, Carton, Muir, Ghosh, Downey, Echlin, Pollock, and Daly]{brodnik2023perspective}
Neal~R Brodnik, Samuel Carton, Caelin Muir, Satanu Ghosh, Doug Downey, McLean~P Echlin, Tresa~M Pollock, and Samantha Daly.
\newblock Perspective: Large language models in applied mechanics.
\newblock \emph{Journal of Applied Mechanics}, 90\penalty0 (10):\penalty0 101008, 2023.

\bibitem[Chang et~al.(2025)Chang, Azvar, Okwudire, and Kontar]{chang_textttllinbo_2025}
Chih-Yu Chang, Milad Azvar, Chinedum Okwudire, and Raed~Al Kontar.
\newblock \${\textbackslash}texttt\{{LLINBO}\}\$: {Trustworthy} {LLM}-in-the-{Loop} {Bayesian} {Optimization}, May 2025.
\newblock URL \url{http://arxiv.org/abs/2505.14756}.
\newblock arXiv:2505.14756 [cs] version: 1.

\bibitem[Chen et~al.(2019)Chen, Ye, Zuo, Zheng, and Ong]{chen2019graph}
Chi Chen, Weike Ye, Yunxing Zuo, Chen Zheng, and Shyue~Ping Ong.
\newblock Graph networks as a universal machine learning framework for molecules and crystals.
\newblock \emph{Chemistry of Materials}, 31\penalty0 (9):\penalty0 3564--3572, 2019.

\bibitem[Chen et~al.(2023)Chen, Wulamu, Zou, Zheng, Wen, Guo, Chen, Zhang, and Zhang]{chen2023md}
Saian Chen, Aziguli Wulamu, Qiping Zou, Han Zheng, Li~Wen, Xi~Guo, Han Chen, Taohong Zhang, and Ying Zhang.
\newblock Md-gnn: A mechanism-data-driven graph neural network for molecular properties prediction and new material discovery.
\newblock \emph{Journal of Molecular Graphics and Modelling}, 123:\penalty0 108506, 2023.

\bibitem[Curtarolo et~al.(2012)Curtarolo, Setyawan, Hart, Jahnatek, Chepulskii, Taylor, Wang, Xue, Yang, Levy, et~al.]{AFLOW_curtarolo2012aflow}
Stefano Curtarolo, Wahyu Setyawan, Gus~LW Hart, Michal Jahnatek, Roman~V Chepulskii, Richard~H Taylor, Shidong Wang, Junkai Xue, Kesong Yang, Ohad Levy, et~al.
\newblock Aflow: An automatic framework for high-throughput materials discovery.
\newblock \emph{Computational Materials Science}, 58:\penalty0 218--226, 2012.

\bibitem[Davies et~al.(2016)Davies, Butler, Jackson, Morris, Frost, Skelton, and Walsh]{davies2016computational}
Daniel~W Davies, Keith~T Butler, Adam~J Jackson, Andrew Morris, Jarvist~M Frost, Jonathan~M Skelton, and Aron Walsh.
\newblock Computational screening of all stoichiometric inorganic materials.
\newblock \emph{Chem}, 1\penalty0 (4):\penalty0 617--627, 2016.

\bibitem[Frey et~al.(2022)Frey, Silverstein, and Pollock]{BCC_B2_frey2022high}
Carolina Frey, Ravit Silverstein, and Tresa~M Pollock.
\newblock A high stability b2-containing refractory multi-principal element alloy.
\newblock \emph{Acta Materialia}, 229:\penalty0 117767, 2022.

\bibitem[Frey et~al.(2024)Frey, You, Kube, Balbus, Mullin, Oppenheimer, Holgate, and Pollock]{BCC_B2_frey2024high}
Carolina Frey, Haojun You, Sebastian Kube, Glenn~H Balbus, Kaitlyn Mullin, Scott Oppenheimer, Collin~S Holgate, and Tresa~M Pollock.
\newblock High temperature b2 precipitation in ru-containing refractory multi-principal element alloys.
\newblock \emph{Metallurgical and Materials Transactions A}, 55\penalty0 (6):\penalty0 1739--1764, 2024.

\bibitem[Gan et~al.(2025)Gan, Zhong, Du, Zhu, Duan, Wang, Schwalbe-Koda, Gomes, Persson, and Wang]{gan2025large}
Jingru Gan, Peichen Zhong, Yuanqi Du, Yanqiao Zhu, Chenru Duan, Haorui Wang, Daniel Schwalbe-Koda, Carla~P Gomes, Kristin Persson, and Wei Wang.
\newblock Large language models are innate crystal structure generators.
\newblock In \emph{AI for Accelerated Materials Design-ICLR 2025}, 2025.

\bibitem[Geerlings et~al.(2003)Geerlings, De~Proft, and Langenaeker]{DFT_geerlings2003conceptual}
Paul Geerlings, Frank De~Proft, and Wilfried Langenaeker.
\newblock Conceptual density functional theory.
\newblock \emph{Chemical reviews}, 103\penalty0 (5):\penalty0 1793--1874, 2003.

\bibitem[Grattafiori et~al.(2024)Grattafiori, Dubey, Jauhri, Pandey, Kadian, Al-Dahle, Letman, Mathur, Schelten, Vaughan, et~al.]{grattafiori2024LLaMA}
Aaron Grattafiori, Abhimanyu Dubey, Abhinav Jauhri, Abhinav Pandey, Abhishek Kadian, Ahmad Al-Dahle, Aiesha Letman, Akhil Mathur, Alan Schelten, Alex Vaughan, et~al.
\newblock The llama 3 herd of models.
\newblock \emph{arXiv preprint arXiv:2407.21783}, 2024.

\bibitem[Gruver et~al.(2024)Gruver, Sriram, Madotto, Wilson, Zitnick, and Ulissi]{gruver2024fine}
Nate Gruver, Anuroop Sriram, Andrea Madotto, Andrew~Gordon Wilson, C~Lawrence Zitnick, and Zachary Ulissi.
\newblock Fine-tuned language models generate stable inorganic materials as text.
\newblock \emph{arXiv preprint arXiv:2402.04379}, 2024.

\bibitem[Hastings et~al.(2025)Hastings, Mulukutla, Khatamsaz, Salas, Xu, Lewis, Person, Skokan, Miller, Paramore, Butler, Allaire, Attari, Karaman, Pharr, Srivastava, and Arroyave]{hastings_accelerated_2025}
Trevor Hastings, Mrinalini Mulukutla, Danial Khatamsaz, Daniel Salas, Wenle Xu, Daniel Lewis, Nicole Person, Matthew Skokan, Braden Miller, James Paramore, Brady Butler, Douglas Allaire, Vahid Attari, Ibrahim Karaman, George Pharr, Ankit Srivastava, and Raymundo Arroyave.
\newblock Accelerated {Multi}-{Objective} {Alloy} {Discovery} through {Efficient} {Bayesian} {Methods}: {Application} to the {FCC} {Alloy} {Space}, March 2025.
\newblock URL \url{http://arxiv.org/abs/2405.08900}.
\newblock arXiv:2405.08900 [cond-mat].

\bibitem[Hobson(1962)]{refractories_osti_4515686}
D.~O. Hobson.
\newblock Effect of alloying elements on the strength, stability, and corrosion and oxidation resistance of columbium. a literature survey.
\newblock Technical report, Oak Ridge National Lab. (ORNL), Oak Ridge, TN (United States), 03 1962.
\newblock URL \url{https://www.osti.gov/biblio/4515686}.

\bibitem[Holgate et~al.(2025)Holgate, Frey, Endsley, Suzuki, Levi, and Pollock]{holgate_design_2025}
Collin~S. Holgate, Carolina Frey, Melina~A. Endsley, Akane Suzuki, Carlos~G. Levi, and Tresa~M. Pollock.
\newblock Design of an alumina forming coating for {Nb}-base refractory alloys.
\newblock \emph{Materials \& Design}, 251:\penalty0 113652, March 2025.
\newblock ISSN 0264-1275.
\newblock \doi{10.1016/j.matdes.2025.113652}.
\newblock URL \url{https://www.sciencedirect.com/science/article/pii/S0264127525000723}.

\bibitem[Horgan(2021)]{horgan_biased_2021}
Madison Horgan.
\newblock Biased decision making in materials science: {Where} does it originate and can it be avoided?
\newblock \emph{MRS Bulletin}, 46\penalty0 (5):\penalty0 361--367, May 2021.
\newblock ISSN 1938-1425.
\newblock \doi{10.1557/s43577-021-00104-5}.
\newblock URL \url{https://doi.org/10.1557/s43577-021-00104-5}.

\bibitem[Hu et~al.(2022)Hu, Shen, Wallis, Allen-Zhu, Li, Wang, Wang, Chen, et~al.]{hu2022lora}
Edward~J Hu, Yelong Shen, Phillip Wallis, Zeyuan Allen-Zhu, Yuanzhi Li, Shean Wang, Lu~Wang, Weizhu Chen, et~al.
\newblock Lora: Low-rank adaptation of large language models.
\newblock \emph{ICLR}, 1\penalty0 (2):\penalty0 3, 2022.

\bibitem[Humphrey et~al.(1996)Humphrey, Dalke, and Schulten]{Visual_MD_humphrey1996vmd}
William Humphrey, Andrew Dalke, and Klaus Schulten.
\newblock Vmd: visual molecular dynamics.
\newblock \emph{Journal of molecular graphics}, 14\penalty0 (1):\penalty0 33--38, 1996.

\bibitem[Jain et~al.(2013)Jain, Ong, Hautier, Chen, Richards, Dacek, Cholia, Gunter, Skinner, Ceder, and Persson]{Jain2013MaterialsProject}
Anubhav Jain, Shyue~Ping Ong, Geoffroy Hautier, Wei Chen, William~Davidson Richards, Stephen Dacek, Shreyas Cholia, Dan Gunter, David Skinner, Gerbrand Ceder, and Kristin~A. Persson.
\newblock Commentary: The materials project: A materials genome approach to accelerating materials innovation.
\newblock \emph{APL Materials}, 1\penalty0 (1):\penalty0 011002, 2013.
\newblock \doi{10.1063/1.4812323}.

\bibitem[Jia et~al.(2019)Jia, Lynch, Huang, Danielson, Lang'at, Milder, Ruby, Wang, Friedler, Norquist, and Schrier]{jia_anthropogenic_2019}
Xiwen Jia, Allyson Lynch, Yuheng Huang, Matthew Danielson, Immaculate Lang'at, Alexander Milder, Aaron~E. Ruby, Hao Wang, Sorelle~A. Friedler, Alexander~J. Norquist, and Joshua Schrier.
\newblock Anthropogenic biases in chemical reaction data hinder exploratory inorganic synthesis.
\newblock \emph{Nature}, 573\penalty0 (7773):\penalty0 251--255, September 2019.
\newblock ISSN 1476-4687.
\newblock \doi{10.1038/s41586-019-1540-5}.

\bibitem[Kaufman and Nesor(1974)]{Calphad_kaufman1974calculation}
Larry Kaufman and Harvey Nesor.
\newblock Calculation of superalloy phase diagrams: Part i.
\newblock \emph{Metallurgical and Materials Transactions B}, 5:\penalty0 1617--1621, 1974.

\bibitem[Khattab et~al.(2023)Khattab, Singhvi, Maheshwari, Zhang, Santhanam, Vardhamanan, Haq, Sharma, Joshi, Moazam, et~al.]{khattab2023dspy}
Omar Khattab, Arnav Singhvi, Paridhi Maheshwari, Zhiyuan Zhang, Keshav Santhanam, Sri Vardhamanan, Saiful Haq, Ashutosh Sharma, Thomas~T Joshi, Hanna Moazam, et~al.
\newblock Dspy: Compiling declarative language model calls into self-improving pipelines.
\newblock \emph{arXiv preprint arXiv:2310.03714}, 2023.

\bibitem[Kohn et~al.(1996)Kohn, Becke, and Parr]{kohn1996density}
Walter Kohn, Axel~D Becke, and Robert~G Parr.
\newblock Density functional theory of electronic structure.
\newblock \emph{The journal of physical chemistry}, 100\penalty0 (31):\penalty0 12974--12980, 1996.

\bibitem[Kresse and Furthm{\"u}ller(1996{\natexlab{a}})]{VASP_kresse1996software}
Georg Kresse and J~Furthm{\"u}ller.
\newblock Software vasp, vienna (1999).
\newblock \emph{Phys. Rev. B}, 54\penalty0 (11):\penalty0 169, 1996{\natexlab{a}}.

\bibitem[Kresse and Furthm{\"u}ller(1996{\natexlab{b}})]{VASP_kresse1996efficiency}
Georg Kresse and J{\"u}rgen Furthm{\"u}ller.
\newblock Efficiency of ab-initio total energy calculations for metals and semiconductors using a plane-wave basis set.
\newblock \emph{Computational materials science}, 6\penalty0 (1):\penalty0 15--50, 1996{\natexlab{b}}.

\bibitem[Kresse and Hafner(1993)]{VASP_kresse1993ab}
Georg Kresse and J{\"u}rgen Hafner.
\newblock Ab initio molecular dynamics for liquid metals.
\newblock \emph{Physical review B}, 47\penalty0 (1):\penalty0 558, 1993.

\bibitem[Kube et~al.(2024{\natexlab{a}})Kube, Frey, McMullin, Neuman, Mullin, and Pollock]{BCC_B2_kube2024navigating}
Sebastian~A Kube, Carolina Frey, Chiyo McMullin, Ben Neuman, Kaitlyn~M Mullin, and Tresa~M Pollock.
\newblock Navigating the bcc-b2 refractory alloy space: Stability and thermal processing with ru-b2 precipitates.
\newblock \emph{Acta Materialia}, 265:\penalty0 119628, 2024{\natexlab{a}}.

\bibitem[Kube et~al.(2024{\natexlab{b}})Kube, Frey, McMullin, and Pollock]{Kube2024BCCB2}
Sebastian~A. Kube, Carolina Frey, Chiyo McMullin, and Tresa~M. Pollock.
\newblock Navigating the bcc–b2 refractory alloy space: Stability and thermal processing with ru–b2 precipitates.
\newblock \emph{Acta Materialia}, 265:\penalty0 119628, 2024{\natexlab{b}}.
\newblock \doi{10.1016/j.actamat.2023.119628}.
\newblock URL \url{https://doi.org/10.1016/j.actamat.2023.119628}.

\bibitem[Lai and Pu(2025)]{lai2025prim}
Ryan~Zheyuan Lai and Yingming Pu.
\newblock Prim: Principle-inspired material discovery through multi-agent collaboration.
\newblock In \emph{AI for Accelerated Materials Design-ICLR 2025}, 2025.

\bibitem[Li et~al.(2020)Li, Li, Wang, Dong, and Liaw]{BCC_B2_li2020phase}
JL~Li, Z~Li, Q~Wang, C~Dong, and PK~Liaw.
\newblock Phase-field simulation of coherent bcc/b2 microstructures in high entropy alloys.
\newblock \emph{Acta Materialia}, 197:\penalty0 10--19, 2020.

\bibitem[Li et~al.(2025)Li, Cao, Jiao, Wang, Wang, Liu, Chen, Li, Liu, Rong, Wang, Zhang, and Yu]{li_materials_2025}
Zhixun Li, Bin Cao, Rui Jiao, Liang Wang, Ding Wang, Yang Liu, Dingshuo Chen, Jia Li, Qiang Liu, Yu~Rong, Liang Wang, Tong-yi Zhang, and Jeffrey~Xu Yu.
\newblock Materials {Generation} in the {Era} of {Artificial} {Intelligence}: {A} {Comprehensive} {Survey}, May 2025.
\newblock URL \url{http://arxiv.org/abs/2505.16379}.
\newblock arXiv:2505.16379 [cond-mat].

\bibitem[Liu et~al.(2024)Liu, Astorga, Seedat, and Schaar]{liu_large_2024}
Tennison Liu, Nicolás Astorga, Nabeel Seedat, and Mihaela van~der Schaar.
\newblock Large {Language} {Models} to {Enhance} {Bayesian} {Optimization}, March 2024.
\newblock URL \url{http://arxiv.org/abs/2402.03921}.
\newblock arXiv:2402.03921 [cs].

\bibitem[Ma et~al.(2017)Ma, Jiang, Li, Wang, Dong, Liaw, Xu, and Sun]{BCC_B2_ma2017bcc}
Yue Ma, Beibei Jiang, Chunling Li, Qing Wang, Chuang Dong, Peter~K Liaw, Fen Xu, and Lixian Sun.
\newblock The bcc/b2 morphologies in al x nicofecr high-entropy alloys.
\newblock \emph{Metals}, 7\penalty0 (2):\penalty0 57, 2017.

\bibitem[Merchant et~al.(2023)Merchant, Batzner, Schoenholz, Aykol, Cheon, and Cubuk]{merchant2023scaling}
Amil Merchant, Simon Batzner, Samuel~S Schoenholz, Muratahan Aykol, Gowoon Cheon, and Ekin~Dogus Cubuk.
\newblock Scaling deep learning for materials discovery.
\newblock \emph{Nature}, 624\penalty0 (7990):\penalty0 80--85, 2023.

\bibitem[Naka and Khan(1997)]{refractories_naka1997designing}
S~Naka and T~Khan.
\newblock Designing novel multiconstituent inter metallies: Contribution of modern alloy theory in developing engineered materials.
\newblock \emph{Journal of phase equilibria}, 18\penalty0 (6):\penalty0 635, 1997.

\bibitem[OLMo et~al.(2024)OLMo, Walsh, Soldaini, Groeneveld, Lo, Arora, Bhagia, Gu, Huang, Jordan, et~al.]{olmo20242}
Team OLMo, Pete Walsh, Luca Soldaini, Dirk Groeneveld, Kyle Lo, Shane Arora, Akshita Bhagia, Yuling Gu, Shengyi Huang, Matt Jordan, et~al.
\newblock 2 olmo 2 furious.
\newblock \emph{arXiv preprint arXiv:2501.00656}, 2024.

\bibitem[Rafailov et~al.(2023)Rafailov, Sharma, Mitchell, Manning, Ermon, and Finn]{rafailov2023direct}
Rafael Rafailov, Archit Sharma, Eric Mitchell, Christopher~D Manning, Stefano Ermon, and Chelsea Finn.
\newblock Direct preference optimization: Your language model is secretly a reward model.
\newblock \emph{Advances in Neural Information Processing Systems}, 36:\penalty0 53728--53741, 2023.

\bibitem[Rafailov et~al.(2024)Rafailov, Chittepu, Park, Sikchi, Hejna, Knox, Finn, and Niekum]{rafailov2024scaling}
Rafael Rafailov, Yaswanth Chittepu, Ryan Park, Harshit~Sushil Sikchi, Joey Hejna, Brad Knox, Chelsea Finn, and Scott Niekum.
\newblock Scaling laws for reward model overoptimization in direct alignment algorithms.
\newblock \emph{Advances in Neural Information Processing Systems}, 37:\penalty0 126207--126242, 2024.

\bibitem[Saal et~al.(2013)Saal, Kirklin, Aykol, Meredig, and Wolverton]{OQMD_saal2013materials}
James~E Saal, Scott Kirklin, Muratahan Aykol, Bryce Meredig, and Christopher Wolverton.
\newblock Materials design and discovery with high-throughput density functional theory: the open quantum materials database (oqmd).
\newblock \emph{Jom}, 65:\penalty0 1501--1509, 2013.

\bibitem[Seshadri and Cheetham(2024)]{Seshadri2024GNoMECritique}
Ram Seshadri and Anthony~K. Cheetham.
\newblock Viewpoint: Are the “2.2 million new materials” from gnome really new, and are they materials?
\newblock \emph{Chemistry of Materials}, 36\penalty0 (7):\penalty0 2681--2683, 2024.
\newblock \doi{10.1021/acs.chemmater.4c00643}.
\newblock URL \url{https://pubs.acs.org/doi/10.1021/acs.chemmater.4c00643}.

\bibitem[Shaysultanov et~al.(2017)Shaysultanov, Salishchev, Ivanisenko, Zherebtsov, Tikhonovsky, and Stepanov]{BCC_B2_shaysultanov2017novel}
DG~Shaysultanov, GA~Salishchev, Yu~V Ivanisenko, SV~Zherebtsov, MA~Tikhonovsky, and ND~Stepanov.
\newblock Novel fe36mn21cr18ni15al10 high entropy alloy with bcc/b2 dual-phase structure.
\newblock \emph{Journal of Alloys and Compounds}, 705:\penalty0 756--763, 2017.

\bibitem[Shi et~al.(2025)Shi, Xin, Huo, Jiang, Wu, Chen, Qin, Ma, Huang, Wang, et~al.]{shi2025fine}
Zhuofan Shi, Chunxiao Xin, Tong Huo, Yuntao Jiang, Bowen Wu, Xingyue Chen, Wei Qin, Xinjian Ma, Gang Huang, Zhenyu Wang, et~al.
\newblock A fine-tuned large language model based molecular dynamics agent for code generation to obtain material thermodynamic parameters.
\newblock \emph{Scientific Reports}, 15\penalty0 (1):\penalty0 10295, 2025.

\bibitem[Sriram et~al.(2024)Sriram, Miller, Chen, and Wood]{sriram2024flowllm}
Anuroop Sriram, Benjamin Miller, Ricky~TQ Chen, and Brandon Wood.
\newblock Flowllm: Flow matching for material generation with large language models as base distributions.
\newblock \emph{Advances in Neural Information Processing Systems}, 37:\penalty0 46025--46046, 2024.

\bibitem[Sulzer et~al.(2021)Sulzer, Marquis, Timms, Robinson, and Chapman]{sulzer2021python}
Valentin Sulzer, Scott~G Marquis, Robert Timms, Martin Robinson, and S~Jon Chapman.
\newblock Python battery mathematical modelling (pybamm).
\newblock \emph{Journal of Open Research Software}, 9\penalty0 (1), 2021.

\bibitem[Team et~al.(2024)Team, Riviere, Pathak, Sessa, Hardin, Bhupatiraju, Hussenot, Mesnard, Shahriari, Ram{\'e}, et~al.]{team2024gemma}
Gemma Team, Morgane Riviere, Shreya Pathak, Pier~Giuseppe Sessa, Cassidy Hardin, Surya Bhupatiraju, L{\'e}onard Hussenot, Thomas Mesnard, Bobak Shahriari, Alexandre Ram{\'e}, et~al.
\newblock Gemma 2: Improving open language models at a practical size.
\newblock \emph{arXiv preprint arXiv:2408.00118}, 2024.

\bibitem[{Thermo-Calc Software}((Accessed May 2025))]{tchea7thermocalc}
{Thermo-Calc Software}.
\newblock \emph{TCHEA High Entropy Alloys Database version 7}, (Accessed May 2025).
\newblock URL \url{https://thermocalc.com/products/databases/steel-and-fe-alloys/}.

\bibitem[Vela et~al.(2023)Vela, Khatamsaz, Acemi, Karaman, and Arróyave]{vela_data-augmented_2023}
Brent Vela, Danial Khatamsaz, Cafer Acemi, Ibrahim Karaman, and Raymundo Arróyave.
\newblock Data-augmented modeling for yield strength of refractory high entropy alloys: {A} {Bayesian} approach.
\newblock \emph{Acta Materialia}, 261:\penalty0 119351, December 2023.
\newblock ISSN 1359-6454.
\newblock \doi{10.1016/j.actamat.2023.119351}.
\newblock URL \url{https://www.sciencedirect.com/science/article/pii/S135964542300681X}.

\bibitem[Wang et~al.(2022)Wang, Wu, Li, Wang, Li, and Wang]{BCC_B2_wang2022phase}
Jianbin Wang, Qingfeng Wu, Yue Li, Zhijun Wang, Junjie Li, and Jincheng Wang.
\newblock Phase selection of bcc/b2 phases for the improvement of tensile behaviors in fenicral medium entropy alloy.
\newblock \emph{Journal of Alloys and Compounds}, 916:\penalty0 165382, 2022.

\bibitem[Wang and Dowling(2022)]{wang_bayesian_2022}
Ke~Wang and Alexander~W Dowling.
\newblock Bayesian optimization for chemical products and functional materials.
\newblock \emph{Current Opinion in Chemical Engineering}, 36:\penalty0 100728, June 2022.
\newblock ISSN 2211-3398.
\newblock \doi{10.1016/j.coche.2021.100728}.
\newblock URL \url{https://www.sciencedirect.com/science/article/pii/S2211339821000605}.

\bibitem[Wang et~al.(2018)Wang, Li, Pang, Li, Dong, and Liaw]{precipitates_wang2018coherent}
Qing Wang, Zhen Li, Shujie Pang, Xiaona Li, Chuang Dong, and Peter~K Liaw.
\newblock Coherent precipitation and strengthening in compositionally complex alloys: a review.
\newblock \emph{Entropy}, 20\penalty0 (11):\penalty0 878, 2018.

\bibitem[Ward et~al.(2018)Ward, Dunn, Faghaninia, Zimmermann, Bajaj, Wang, Montoya, Chen, Bystrom, Dylla, et~al.]{ward2018matminer}
Logan Ward, Alexander Dunn, Alireza Faghaninia, Nils~ER Zimmermann, Saurabh Bajaj, Qi~Wang, Joseph Montoya, Jiming Chen, Kyle Bystrom, Maxwell Dylla, et~al.
\newblock Matminer: An open source toolkit for materials data mining.
\newblock \emph{Computational Materials Science}, 152:\penalty0 60--69, 2018.

\bibitem[Whitfield et~al.(2020)Whitfield, Pickering, Owen, Jones, Stone, and Jones]{BCC_B2_whitfield2020effect}
TE~Whitfield, EJ~Pickering, LR~Owen, CN~Jones, HJ~Stone, and NG~Jones.
\newblock The effect of al on the formation and stability of a bcc--b2 microstructure in a refractory metal high entropy superalloy system.
\newblock \emph{Materialia}, 13:\penalty0 100858, 2020.

\bibitem[Xiao et~al.(2023)Xiao, Li, Shi, Chen, Zhu, Chen, and Wang]{xiao2023invertible}
Hang Xiao, Rong Li, Xiaoyang Shi, Yan Chen, Liangliang Zhu, Xi~Chen, and Lei Wang.
\newblock An invertible, invariant crystal representation for inverse design of solid-state materials using generative deep learning.
\newblock \emph{Nature Communications}, 14\penalty0 (1):\penalty0 7027, 2023.

\bibitem[Xie et~al.(2021)Xie, Fu, Ganea, Barzilay, and Jaakkola]{xie2021crystal}
Tian Xie, Xiang Fu, Octavian-Eugen Ganea, Regina Barzilay, and Tommi Jaakkola.
\newblock Crystal diffusion variational autoencoder for periodic material generation.
\newblock \emph{arXiv preprint arXiv:2110.06197}, 2021.

\bibitem[Xu et~al.(2025)Xu, Desai, Wang, Hope, and Ritz]{xu_plaid_2025}
Andy Xu, Rohan Desai, Larry Wang, Gabriel Hope, and Ethan~T. Ritz.
\newblock {PLaID}: {Preference} {Aligned} {Language} {Model} for {Targeted} {Inorganic} {Materials} {Design}.
\newblock April 2025.
\newblock URL \url{https://openreview.net/forum?id=7aoP3ZeBfy}.

\bibitem[Yan et~al.(2024)Yan, Li, Ling, Ashen, Edwards, Arr{\'o}yave, Zitnik, Ji, Qian, Qian, et~al.]{yan2024invariant}
Keqiang Yan, Xiner Li, Hongyi Ling, Kenna Ashen, Carl Edwards, Raymundo Arr{\'o}yave, Marinka Zitnik, Heng Ji, Xiaofeng Qian, Xiaoning Qian, et~al.
\newblock Invariant tokenization of crystalline materials for language model enabled generation.
\newblock \emph{Advances in Neural Information Processing Systems}, 37:\penalty0 125050--125072, 2024.

\bibitem[Yang et~al.(2024)Yang, Batzner, Gao, Aykol, Gaunt, McMorrow, Jimenez~Rezende, Schuurmans, Mordatch, and Cubuk]{yang2024generative}
Sherry Yang, Simon Batzner, Ruiqi Gao, Muratahan Aykol, Alexander Gaunt, Brendan~C McMorrow, Danilo Jimenez~Rezende, Dale Schuurmans, Igor Mordatch, and Ekin~Dogus Cubuk.
\newblock Generative hierarchical materials search.
\newblock \emph{Advances in Neural Information Processing Systems}, 37:\penalty0 38799--38819, 2024.

\bibitem[Yurchenko et~al.(2021)Yurchenko, Panina, Shaysultanov, Zherebtsov, and Stepanov]{precipitates_yurchenko2021refractory}
Nikita Yurchenko, Evgeniya Panina, Dmitry Shaysultanov, Sergey Zherebtsov, and Nikita Stepanov.
\newblock Refractory high entropy alloy with ductile intermetallic b2 matrix/hard bcc particles and exceptional strain hardening capacity.
\newblock \emph{Materialia}, 20:\penalty0 101225, 2021.

\bibitem[Zagorac et~al.(2019)Zagorac, M{\"{u}}ller, Ruehl, Zagorac, and Rehme]{Zagorac:in5024}
D.~Zagorac, H.~M{\"{u}}ller, S.~Ruehl, J.~Zagorac, and S.~Rehme.
\newblock {Recent developments in the Inorganic Crystal Structure Database: theoretical crystal structure data and related features}.
\newblock \emph{Journal of Applied Crystallography}, 52\penalty0 (5):\penalty0 918--925, Oct 2019.
\newblock \doi{10.1107/S160057671900997X}.
\newblock URL \url{https://doi.org/10.1107/S160057671900997X}.

\end{thebibliography}
\bibliographystyle{plainnat}

%%%%%%%%%%%%%%%%%%%%%%%%%%%%%%%%%%%%%%%%%%%%%%%%%%%%%%%%%%%%

\clearpage
\appendix
\appendix
\section{Appendix}
We add more technical detail and approach of our work in here.

\subsection{Baselines}
\label{baseline}
Baseline approaches include (1) random search, (2) static prompting of API-based models, (3) automatic prompt tuning, and (4) a basic agentic setup. 

\subsubsection{Random Search}
\label{random_search}
Conventional alloy discovery approaches often do parametric sweeps of composition space for promising candidates. We approximate this approach by constructing a grid of BCC- and B2-forming elements and sampling random compositions from it. The BCC and B2 compositions were constructed separately by randomly sampling from the grid of constituent elements, and volume percentages individually with heuristic rules enforcing likely BCC- and B2-formation, e.g. that B2s must be a 1-to-1 ratio of two B2-forming elements. This method better imitates a traditional parametric search in conventional alloy discovery than randomly sampling known BCC/B2 pairs, as is done in the preparation of the SFT training data.

\subsubsection{Prompting}
\label{API_prompt}
Our second baseline consists of one-shot and few-shot prompting of three state-of-the-art proprietary API-based models: Gemini-2.5, GPT-4.1 and GPT-o3. We find one-shot prompting from these models to be both dominated by few-shot prompting and unreliable in producing valid output formatting, so we do not report results from the former.
In the zero-shot setting, we randomly sample a single exemplar from the SFT model output. 
In the few-shot setting, we provide top 10 and bottom 10 generations from the SFT model as exemplars, ranked on reward.

\begin{figure*}[h]
    \centering
    \includegraphics[scale=0.8]{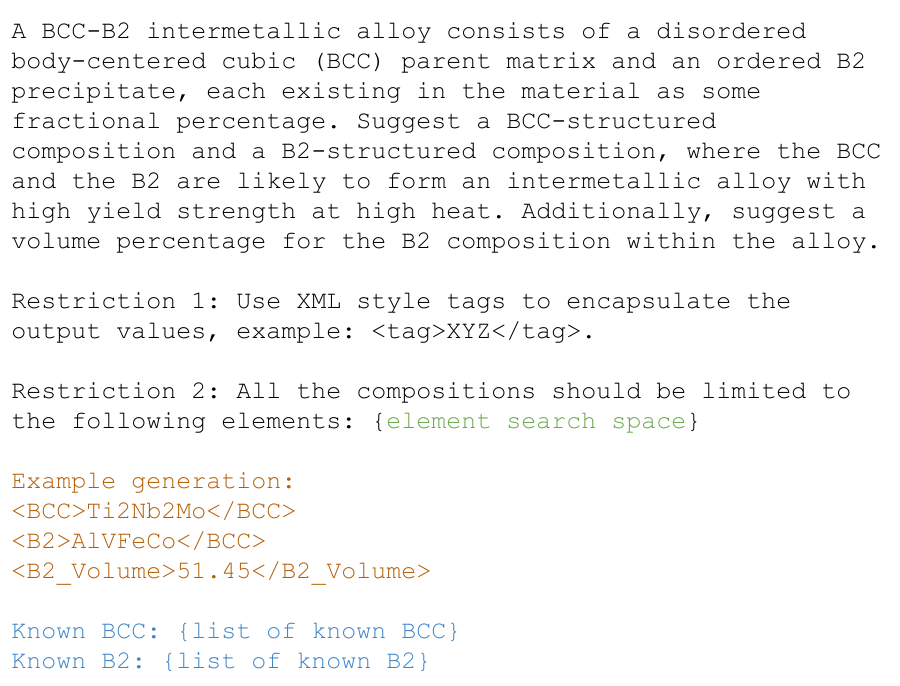}
    \caption{This is the one-shot prompt we used for our API based models. We added some additional context while keeping the training prompt similar. The example generation was randomly sampled from our training data. The text in blue is optional.}
    \label{fig:one_shot_prompt}
\end{figure*}

\begin{figure*}[h]
    \centering
    \includegraphics[scale=0.8]{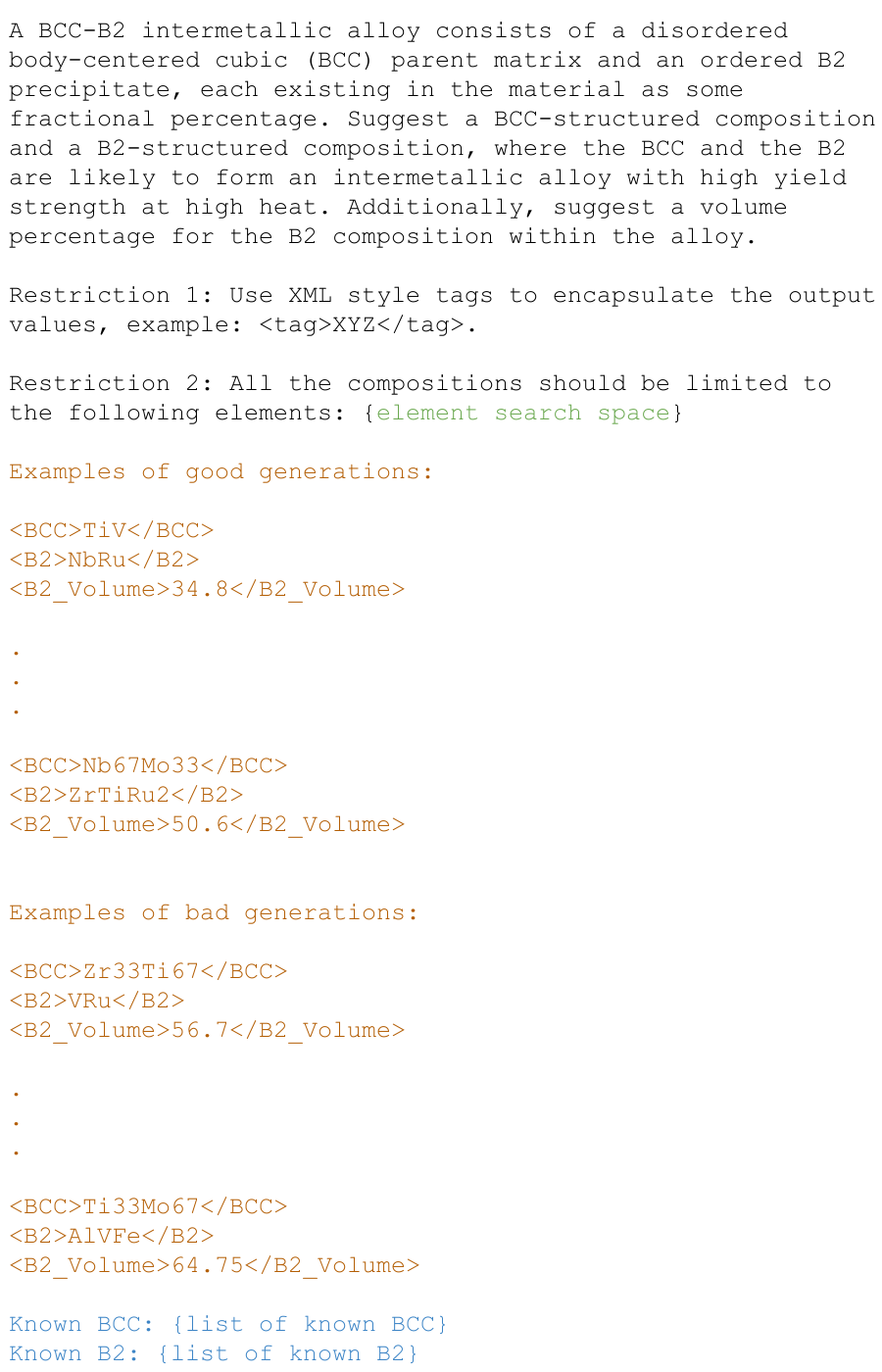}
    \caption{This is the few-shot prompt we used for our API based models. We added some additional context while keeping the training prompt similar. Top-10 and bottom-10 of LLaMA SFT model generations were given here as examples of good and bad generations respectively (only two are shown here for brevity). The text in blue is optional.}
    \label{fig:few_shot_prompt}
\end{figure*}

The prompts that we use for one-shot and few-shot prompting
% in-context tuning of GPT-4.1, GPT-o3, and Gemini-2.5
are provided in Figure~\ref{fig:one_shot_prompt} and Figure~\ref{fig:few_shot_prompt}, respectively. The zero-shot prompting did not work because the models were unable to generate any feasible BCC-B2 pairs in a parseable format.

\subsubsection{Prompt Tuning}
\label{prompt_tuning}
Few-shot Gemini-2.5 produced the most favorable balance of diversity and reward amongst the prompting baselines. To see if this tradeoff could be further optimized, we  create a prompt-tuned few-shot baseline using DSPy\citep{khattab2023dspy}, optimizing the prompt to produce diverse outputs. We used the MIPROv2 with ``medium''-level optimization, using total number of TCHEA unique elements in the alloy system as the metric to optimize. During inference even when using a temperature of 1.0 we could not generate unique triplets with the tuned prompt, which shows robustness of the approach and good for a lot of things but not for us. To sample different composition triplets we added a Universally Unique Identifier (UUID) at the end of each prompt.
As reported in table \ref{tab:basic_results}, this approach does improve diversity at the cost of mean reward.

\subsubsection{Agentic Setup}
\label{agentic_setup}
\begin{figure*}[h]
    \centering
    \includegraphics[scale=0.5]{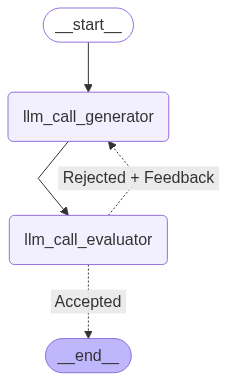}
    \caption{Simple agentic setup with LangChain. We keep the loop going for a maximum of five iterations.}
    \label{fig:agentic_baseline}
\end{figure*}

We implement a simple agentic baseline consisting of a generator and evaluator, implemented with LangGraph (\url{https://github.com/langchain-ai/langgraph}). This is a simple setup where the generator agent is instructed to generate a \triple with few-shot prompt. The generated triple is then sent to the evaluator agent which grades it as ``Valid'' or ``Invalid'' generation, and also provides a detailed reason when judging invalid. If the generation is invalid then we re-route the feedback from the evaluator agent and ask the generator to re-generate the composition. We keep optimizing the generator when it produces invalid \triple for a maximum of five iterations. If the fifth generation is also invalid according to the evaluator we scrap the generation and restart the loop. Otherwise we add it to our acceptable alloy list (as in Figure~\ref{fig:agentic_baseline}). We run this generation-evaluation agentic loop independently until we get 1000 successful generations.

This setup is commonly known as Evaluator-Optimizer~\footnote{\url{https://www.anthropic.com/engineering/building-effective-agents}}. Commonly this setup is used to produce high quality output, as it optimizes the output through iterative refinement~\footnote{\url{https://github.com/OmarKhaled0K/Agents_and_workflows?tab=readme-ov-file}}. In our use-case we wanted the model to produce high quality alloys and hence this setup made most sense (disregarding cost). A similar setup is used by \citet{shi2025fine} to automate generation and refinement of simulation code for materials synthesis.

% This setup is inspired by TKTK, which, while more elaborate in structure, uses a similar generate-validate approach to generate candidate TKTK. 

\subsection{SFT Training Data Curation}
\label{appendix_sft_data}
To build our initial dataset of 207 body-centered cubic (BCC) and 88 B2-structured compositions, a list of known known BCC and B2 structures from the Materials Project \citep{Jain2013MaterialsProject}, was filtered to keep only compounds comprised of the 26 elements in Thermo-Calc’s TCHEA7 database \citep{tchea7thermocalc}. A second filter was then applied to keep only compounds with a calculated energy above the convex hull between 0 and 0.25 eV/atom. (A compound with an energy of 0 eV/atom is expected to be stable at 0 K; by 0.25 eV/atom, a compound is highly unlikely to be stable at 0 K but could become stabilized by entropy effects at elevated temperatures relevant to BCC/B2 alloys.) This processing yielded 24 BCCs (primarily single-element entries) and 57 B2s (exclusively two-element pairs). 

These lists served as the basis for further iteration. First, the role of all elements was estimated. For example, it was noted that elements like Nb and Mo generally formed stable BCCs, whereas Ti and Zr had larger energies above the convex hull and only form BCC structures at elevated temperatures. Likewise, for the B2 compounds, it was noted that elements like Al and Hf generally occupied the A-site, whereas Fe and Ru generally occupied the B-site; some elements, like Mn or V, could occupy either site, whereas others (e.g., Nb or Ta) were found in higher energy (less stable) B2s. These trends were used to iterate BCC compositions with element concentrations of 20\%, 25\%, 33\%, 40\%, 50\%, 67\%, or 75\%; B2 compositions were iterated with 1–2 elements per site (at 25\% or 50\% concentration). A mixture of stable and metastable elements was used throughout this iteration process to ensure a broad representation of potentially stable phases. This process resulted in 2,413 potential BCC compositions and 1,101 potential B2 compositions. Each potential composition was evaluated with Thermo-Calc, and only compositions forming >99\% BCC or B2 were kept, leaving 207 BCC and 88 B2-structured compositions used for SFT. Finally, a volume fraction of B2 intermetallic was prescribed by drawing from existing BCC-B2 alloys and domain expertise. We sampled the B2 volume percentage uniformly within the [20\%, 70\%] interval. Therefore, the supervised dataset consists of structured triplets of the form {BCC, B2, B2 volume proportion}. For each unique BCC–B2 pair, we sampled three distinct volume fractions, resulting in approximately 55,000 triplets. This dataset defines the compositional search space over which our language model operates.

% This is start of another section
\subsection{SFT Training and Validation}
\label{appendix_sft}
Training was conducted with a batch size of 2 across three NVIDIA A40 GPUs with gradient accumulation every 4 steps. Finetuning was performed with 8-bit quantization and low-rank adapters ($rank = 8~and~\alpha=32$)  using the PEFT library~\footnote{\url{https://github.com/huggingface/peft}}. The adapters were only added for ``q\textunderscore proj'' and ``v\textunderscore proj'', this yields maximum learning without parametric overhead~\citep{hu2022lora}. Cosine annealing was used as a learning rate scheduler. The entire training process required about ~93 hours.

\begin{figure*}[h!]
    \centering
    \begin{subfigure}[b]{0.32\linewidth}
        \centering
        \includegraphics[width=\linewidth]{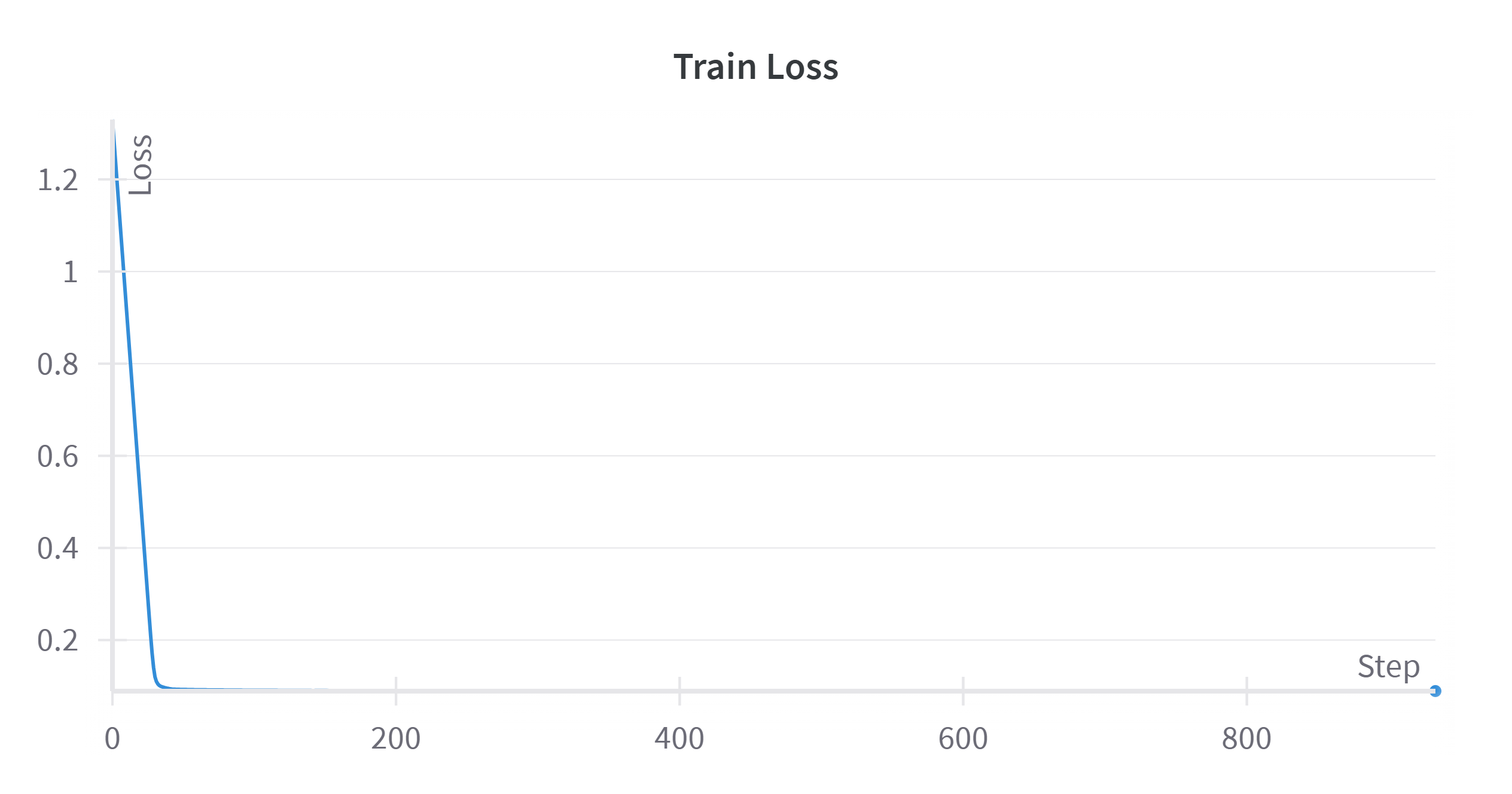}
        \caption{LLaMA SFT Train Loss}
        \label{fig:LLaMA_train_loss}
    \end{subfigure}
    \hfill
    \begin{subfigure}[b]{0.32\linewidth}
        \centering
        \includegraphics[width=\linewidth]{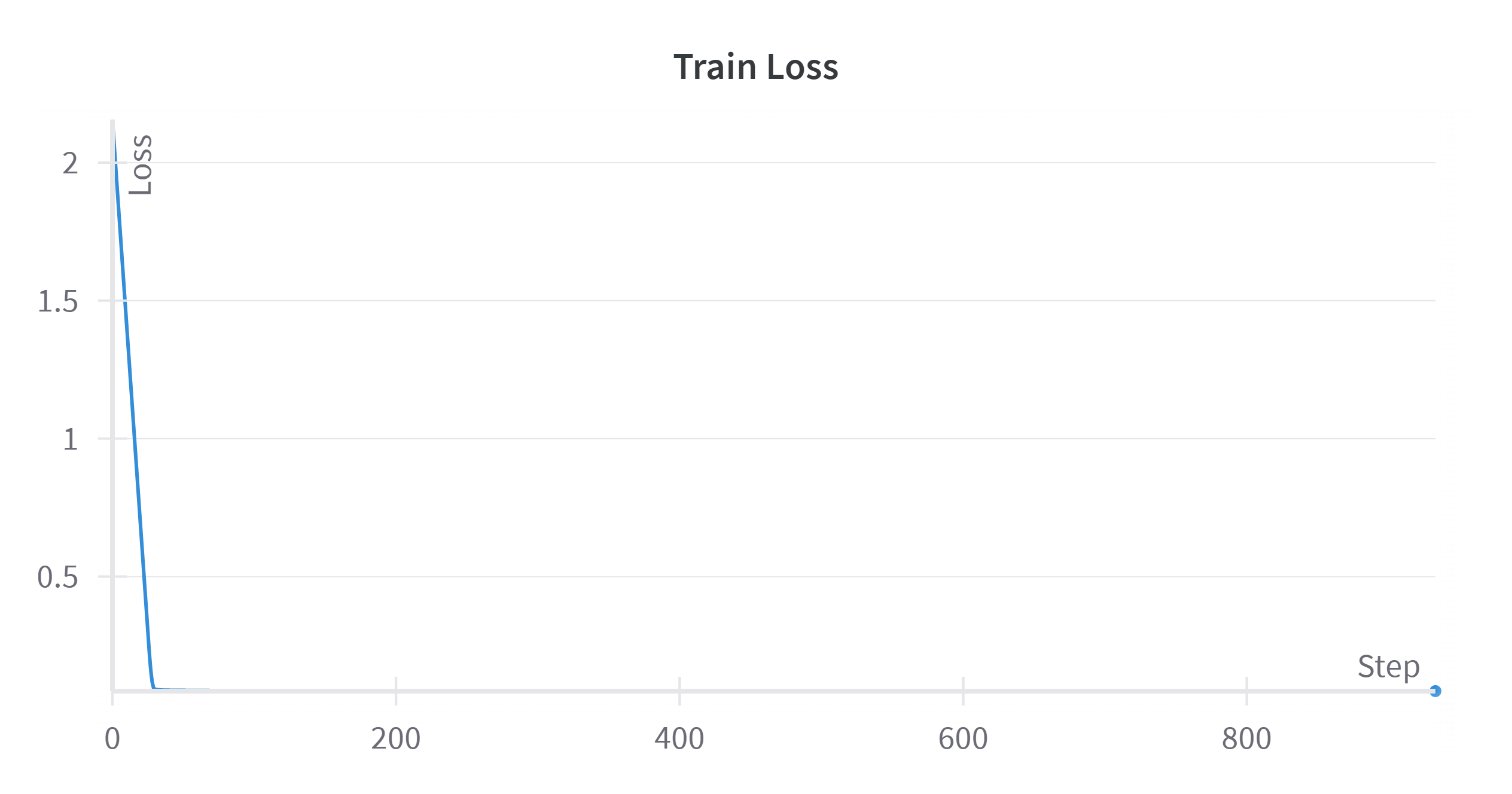}
        \caption{OLMo SFT Train Loss}
        \label{fig:olmo_train_loss}
    \end{subfigure}
    \hfill
    \begin{subfigure}[b]{0.32\linewidth}
        \centering
        \includegraphics[width=\linewidth]{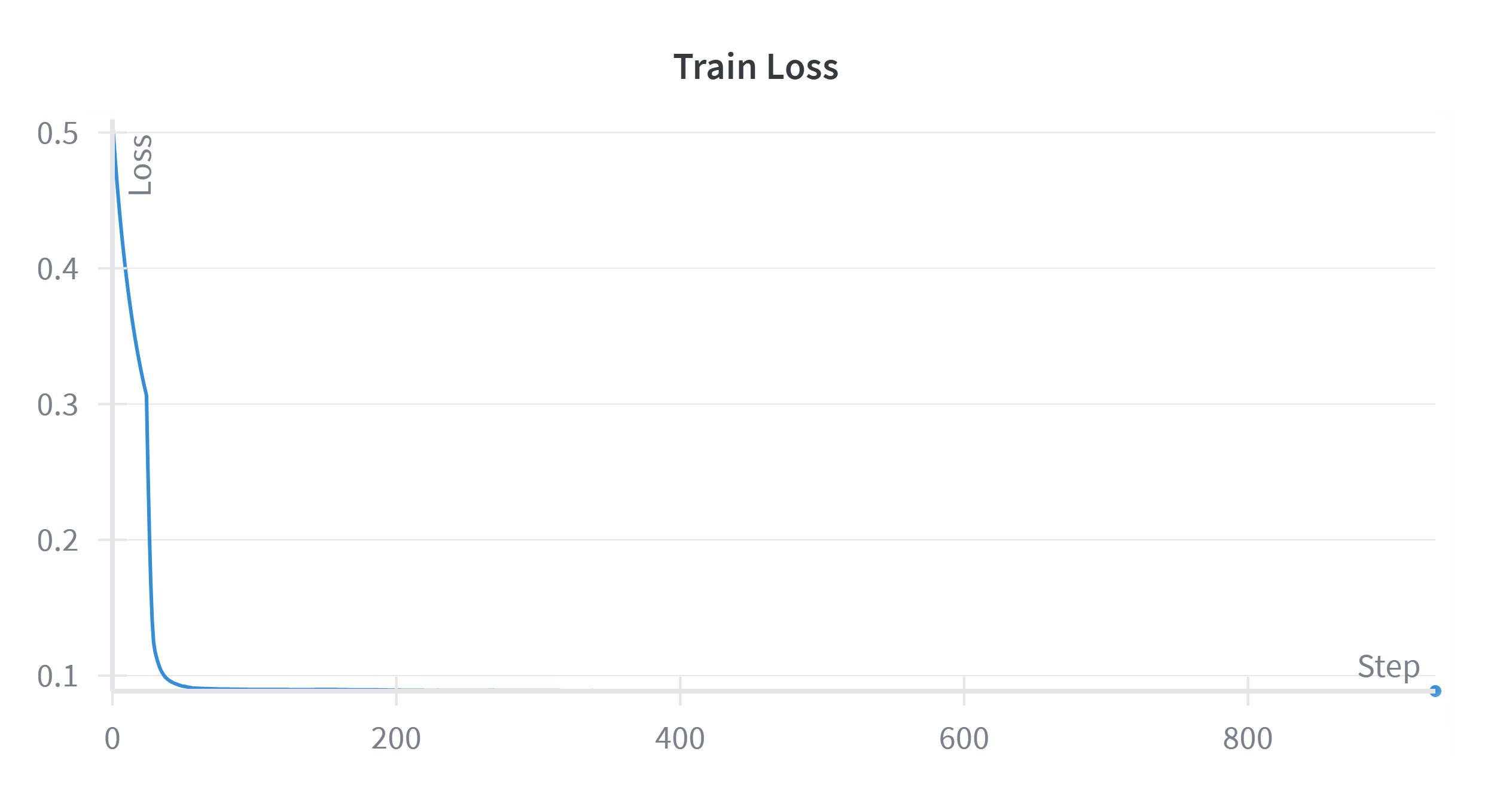}
        \caption{Gemma SFT Train Loss}
        \label{fig:gemma_train_loss}
    \end{subfigure}
    \vfill
    \centering
    \begin{subfigure}[b]{0.32\linewidth}
        \centering
        \includegraphics[width=\linewidth]{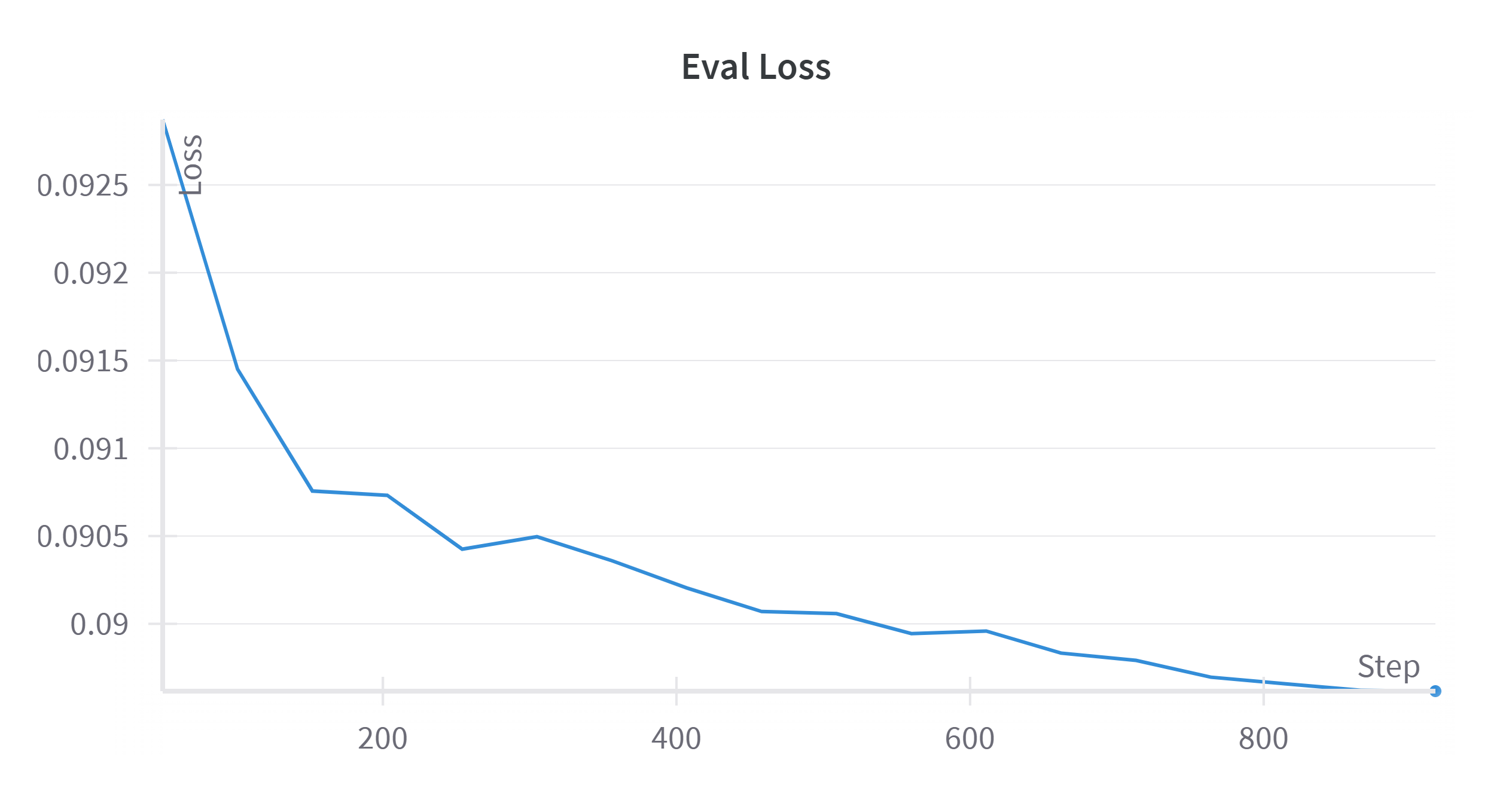}
        \caption{LLaMA SFT Eval Loss}
        \label{fig:LLaMA_eval_loss}
    \end{subfigure}
    \hfill
    \begin{subfigure}[b]{0.32\linewidth}
        \centering
        \includegraphics[width=\linewidth]{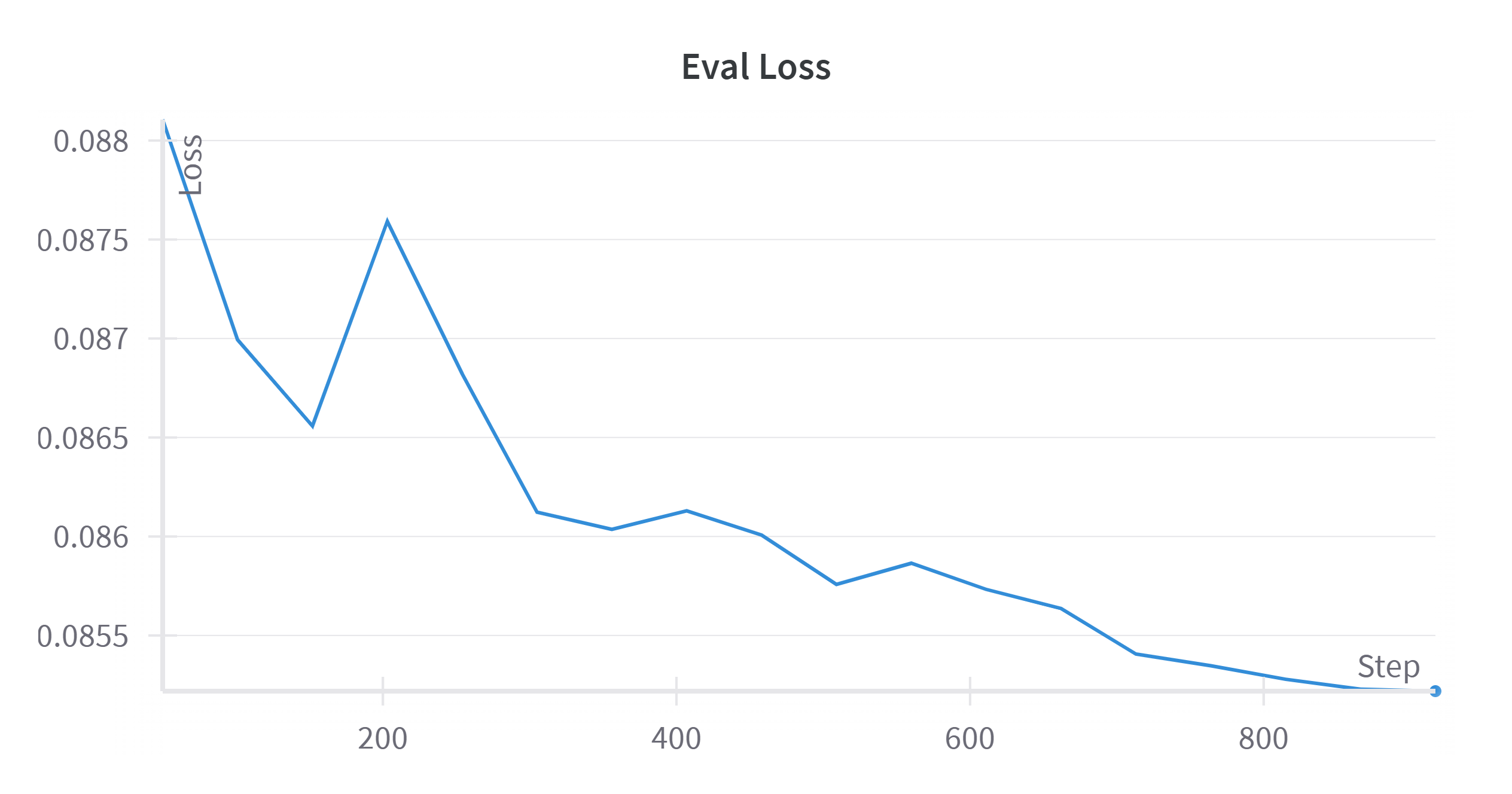}
        \caption{OLMo SFT Eval Loss}
        \label{fig:olmo_eval_loss}
    \end{subfigure}
    \hfill
    \begin{subfigure}[b]{0.32\linewidth}
        \centering
        \includegraphics[width=\linewidth]{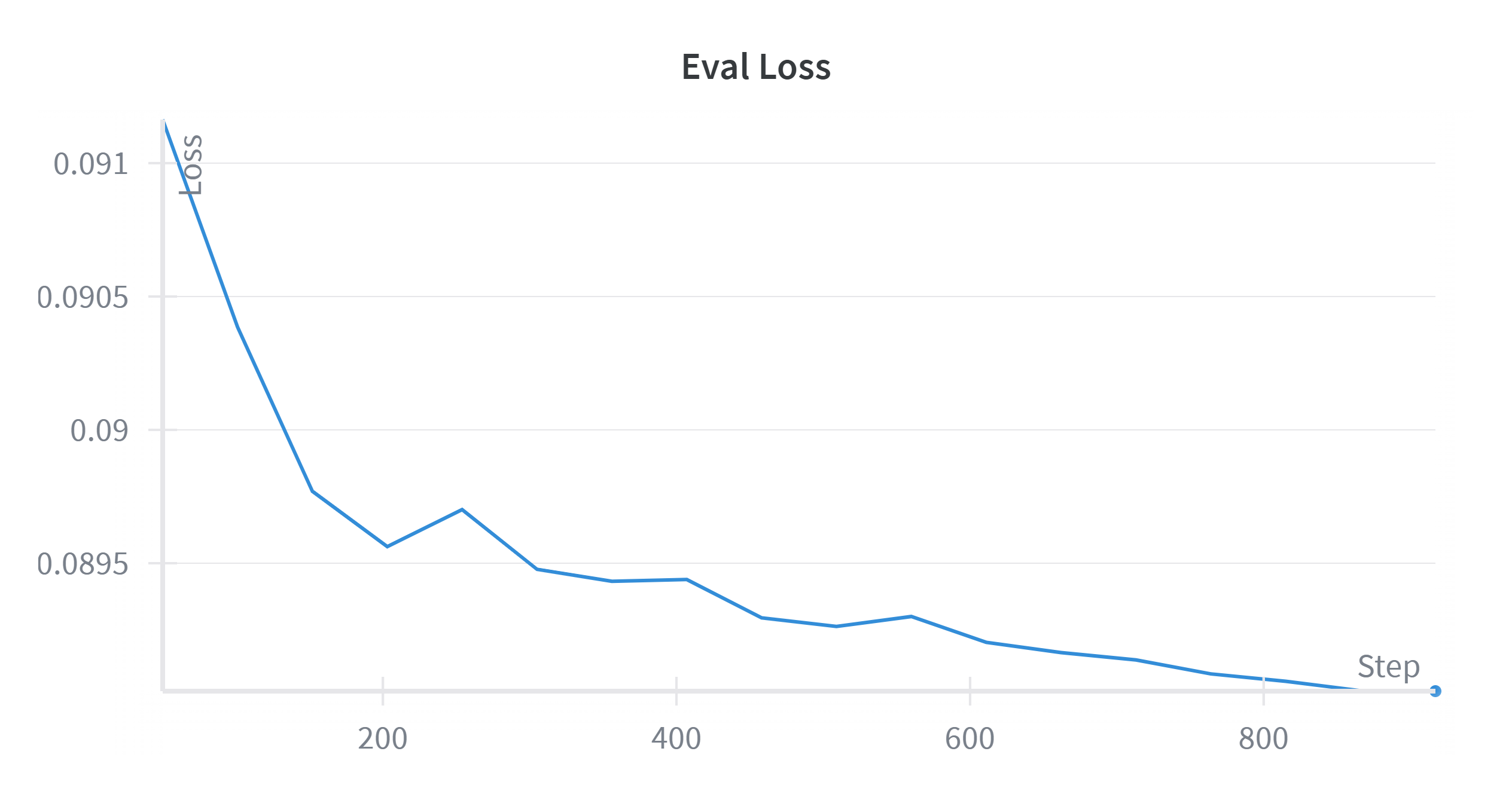}
        \caption{Gemma SFT Eval Loss}
        \label{fig:gemma_eval_loss}
    \end{subfigure}
    \caption{Loss curves for LLaMA, OLMo and Gemma during supervised fine-tuning (SFT).}
    \label{fig:sft_loss}
    
\end{figure*}

The training and evaluation performance for all three local models were similar, as show by loss curves ( Figure~\ref{fig:sft_loss}). Other than a higher starting point for OLMo, the loss curves are almost identical and converge quickly.

We trained each model for 5 epochs. While training loss plateaued after the first epoch, all three models showed steadily declining validation set loss until the end of training. The behavior of OLMo was more unstable than LLaMA or Gemma, but all models converged to a similar validation loss. The 

\label{appendix_dpo}
\subsection{DPO Training Data Curation}
\begin{figure*}[!htpb]
    \centering
    \includegraphics[width=0.95\linewidth]{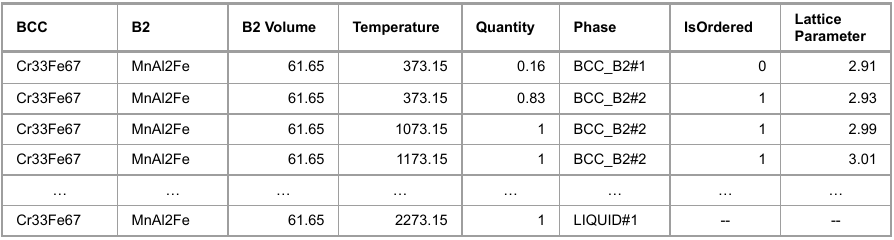}
    \caption{Output from Thermo-Calc evaluates the stability of the generated BCC-B2 alloy over a range of temperatures. The reward function use this output to compute a scalar reward for preference tuning.}
    \label{fig:thermocalc_op}
\end{figure*}

We sample 5000 \triple triples from each SFT model and evaluate them with Thermocalc. Thermocalc predicts the phases of an alloy master composition at different temperatures, resulting in a table where each row represents the predicted portion of a particular phase at a particular temperature (Figure \ref{fig:thermocalc_op}).

We use this feedback to define a reward score for each composition (Eq.~\ref{eq:reward}). The SFT-generated triples are then ranked in descending order by their reward. From this list, we select the top 25\% (1250 triples), each designated as a chosen generation at index $i$. For every chosen generation, we randomly sample 100 distinct rejected generations from index positions with lower reward ($j > i$). This procedure yields $1250 \times 100 = 125{,}000$ preference pairs for DPO training.

We use this particular sampling strategy to balance quality against generalization. Variants we could have explored include a narrower definition of high-quality (e.g. top 10\%), or pairing high-quality candidates against lower-quality candidates (e.g. worst 25\%). We leave it to future work to optimize the sampling strategy for this type of approach.  

\subsection{DPO: Training and Validation}

For DPO, we take the adapter optimized with SFT and perform direct preference optimization. We train on the same configuration as SFT since this was our computational upper limit. We train each model for only 1 epoch. The DPO training took ~70 hours to complete. 

DPO optimizes the following objective:

\begin{align}
& \theta^* = \arg\min_{\theta} \sum_{(x, y^+, y^-) \in \mathcal{D}_{\text{DPO}}} \label{dpo_objective}
 \\
& - \log \sigma \notag
\Big( 
\beta \log \tfrac{\theta(y^+ \mid x)}{\theta_{\text{SFT}}(y^+ \mid x)} 
% \notag\\ 
-\beta \log \tfrac{\theta(y^- \mid x)}{\theta_{\text{SFT}}(y^- \mid x)} 
\Big) 
\end{align}

% \begin{equation}
% \scalebox{0.85}{$
% \begin{split}
% \theta^* = \arg\min_{\theta} \sum_{(x, y^+, y^-) \in \mathcal{D}_{\text{DPO}}} 
% - \log \sigma \Big( 
% \beta \log \tfrac{\theta(y^+ \mid x)}{\theta_{\text{SFT}}(y^+ \mid x)} \\
% - \beta \log \tfrac{\theta(y^- \mid x)}{\theta_{\text{SFT}}(y^- \mid x)} 
% \Big)
% \end{split}
% $}
% \label{dpo_objective}
% \end{equation}

$\theta_\text{SFT}$ and $\theta^*$ are model parameters of SFT and DPO models respectively, $\beta$ is the alternative to KL-penalty factor~\citep{rafailov2023direct}, which controls the distance between the distribution of the $\theta_\text{SFT}$ and $\theta^*$.
We want the internal reward mapping of the model (as no separate reward model is required in DPO) to learn from our multiobjective reward scores and push the model to search the parametric space of higher average reward. However, to prevent the preference tuned model from going wildly out of distribution or hacking the reward function~\citep{rafailov2024scaling}, we set $\beta = 0.5$.

\begin{figure*}[h!]
    \centering
    \begin{subfigure}[b]{0.32\linewidth}
        \centering
        \includegraphics[width=\linewidth]{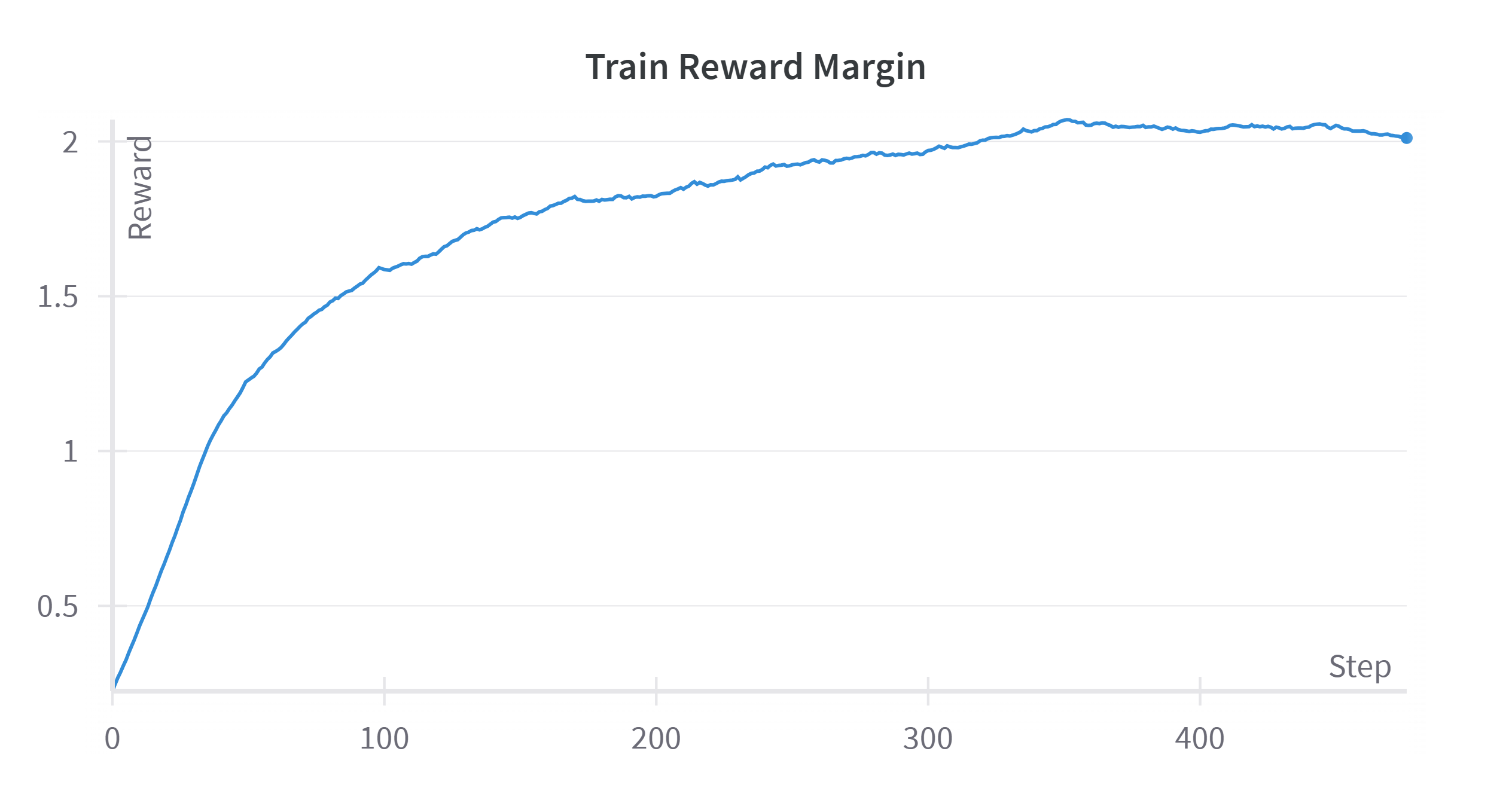}
        \caption{LLaMA DPO Train Reward Margin}
        \label{fig:LLaMA_train_rew_marg}
    \end{subfigure}
    \hfill
    \begin{subfigure}[b]{0.32\linewidth}
        \centering
        \includegraphics[width=\linewidth]{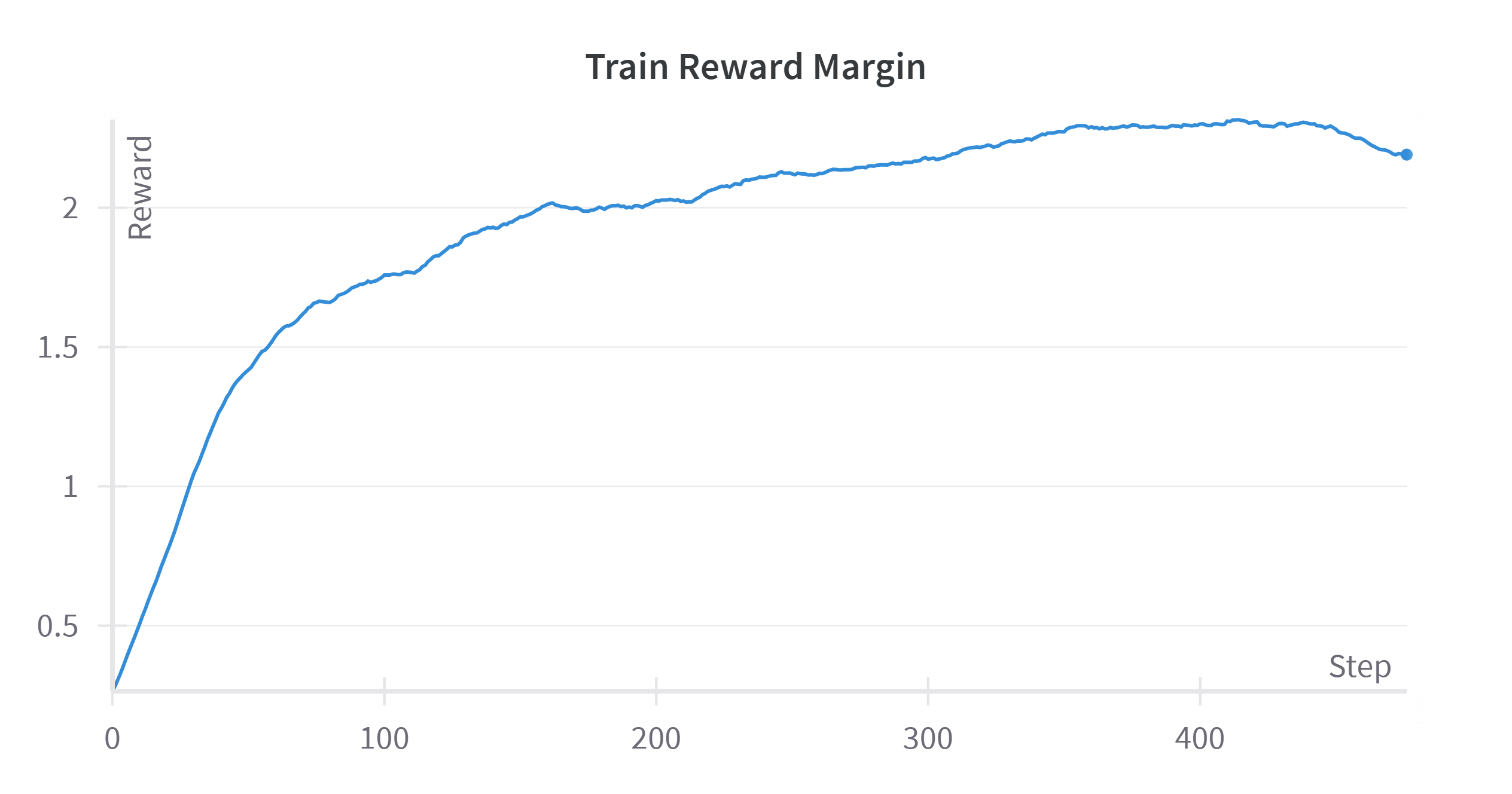}
        \caption{OLMo DPO Train Reward Margin}
        \label{fig:olmo_train_rew_marg}
    \end{subfigure}
    \hfill
    \begin{subfigure}[b]{0.32\linewidth}
        \centering
        \includegraphics[width=\linewidth]{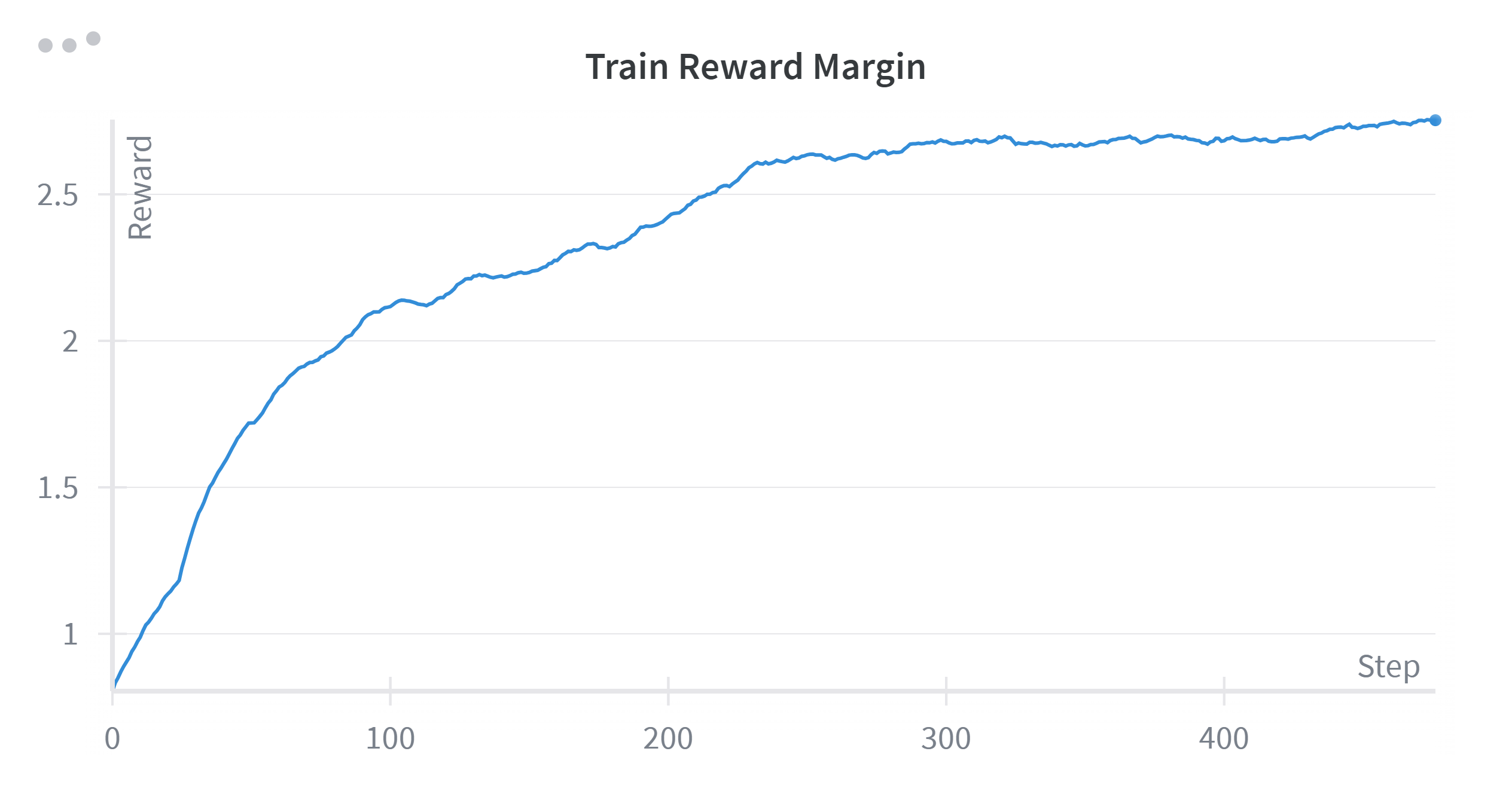}
        \caption{Gemma DPO Train Reward Margin}
        \label{fig:gemma_train_rew_marg}
    \end{subfigure}
    
    \vfill
    \begin{subfigure}[b]{0.32\linewidth}
        \centering
        \includegraphics[width=\linewidth]{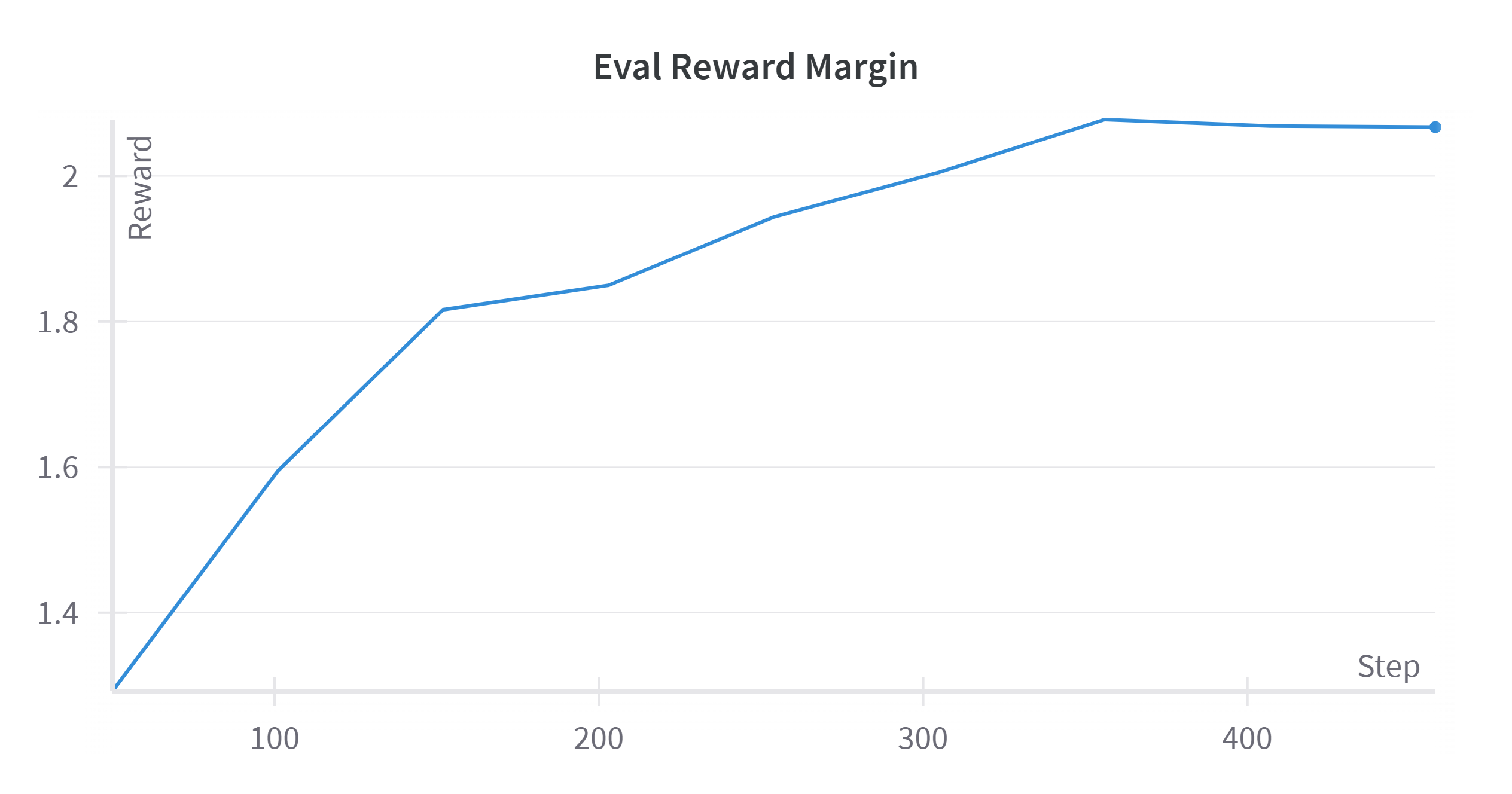}
        \caption{LLaMA DPO Eval Reward Margin}
        \label{fig:LLaMA_eval_rew_marg}
    \end{subfigure}
    \hfill
    \begin{subfigure}[b]{0.32\linewidth}
        \centering
        \includegraphics[width=\linewidth]{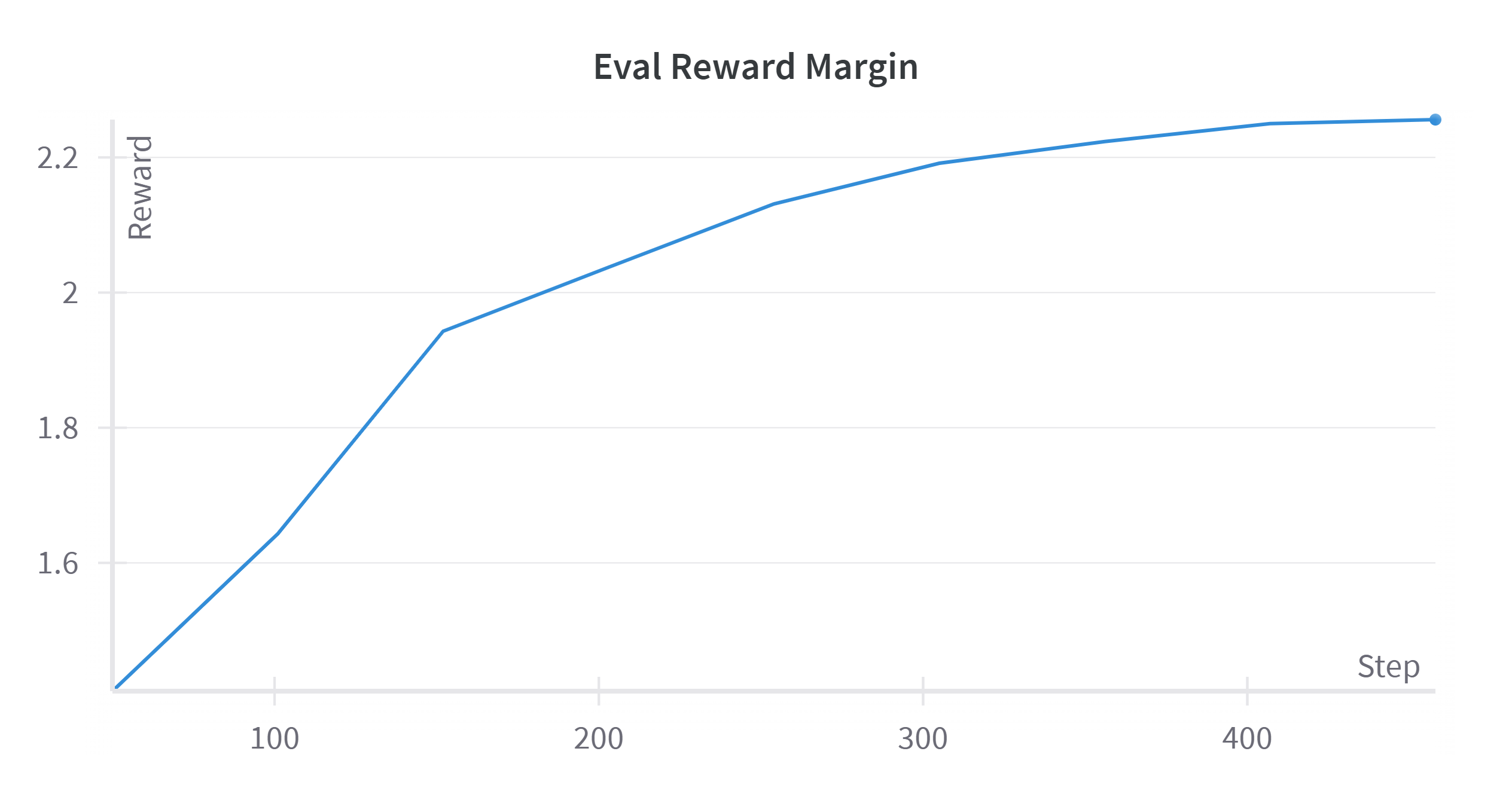}
        \caption{OLMo DPO Eval Reward Margin}
        \label{fig:olmo_eval_rew_marg}
    \end{subfigure}
    \hfill
    \begin{subfigure}[b]{0.32\linewidth}
        \centering
        \includegraphics[width=\linewidth]{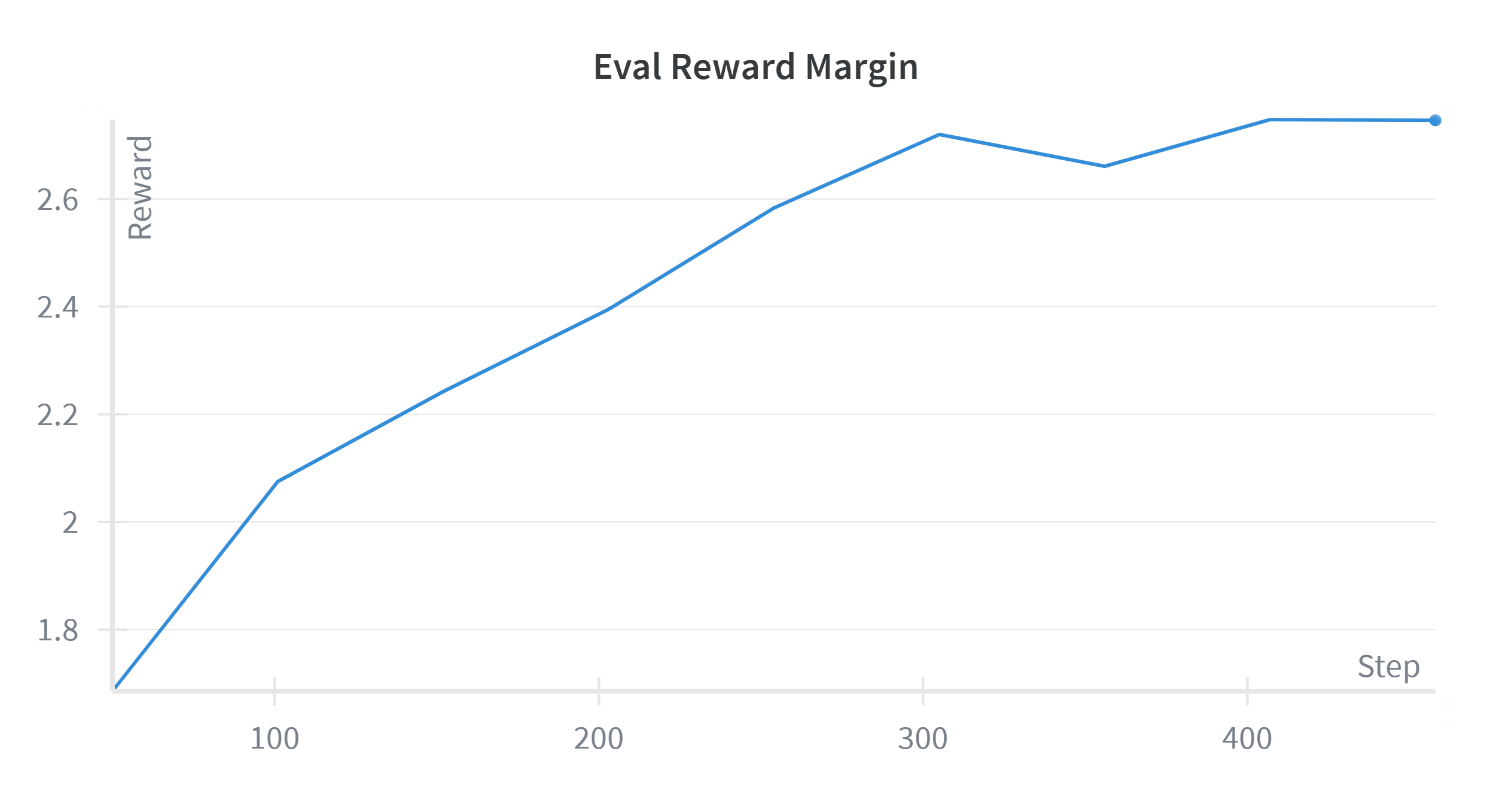}
        \caption{Gemma DPO Eval Reward Margin}
        \label{fig:gemma_eval_rew_marg}
    \end{subfigure}
    \caption{Reward Margin for LLaMA, OLMo and Gemma during preference optimization (DPO).}
    \label{fig:reward_margin}
    
\end{figure*}

The results from all models were again quite similar, with OLMo outperforming LLaMA in terms of reward margin on the evaluation set (Figure~\ref{fig:reward_margin}). We are unsure why OLMo failed to generate higher quality BCC/B2 compositions in spite of its better performance on the evaluation set.

\subsection{Why Preference Tuning Failed on OLMo?}
\label{sec: olmo_fail}
\begin{table*}[h]
	\centering
	\small
	\begin{tabular}{lrrrr}
		\toprule
		\textbf{Element} & \textbf{Count (OLMo)} & \textbf{KL (OLMo)} & \textbf{KL (LLaMA)} & \textbf{KL (Gemma)} \\
		\midrule
		Ti & 94  & 0.0155 & 0.0078 & 4.25e--04 \\
		Al & 53  & 0.0169 & 0.0104 & 3.34e--04 \\
		V  & 49  & 0.0122 & 0.0071 & 1.27e--04 \\
		Nb & 38  & 0.0193 & 0.0030 & 3.33e--04 \\
		W  & 22  & 0.0015 & 0.00015 & 2.98e--05 \\
		Cr & 13  & 0.0289 & 0.0257 & 1.32e--04 \\
		\bottomrule
	\end{tabular}
	\caption{Forward $D_{\mathrm{KL}}(\mathrm{DPO}\,\|\,\mathrm{SFT})$ on generated tokens (teacher--forced; trimmed at EOS) for the elements most frequently produced by OLMo. OLMo’s KL is consistently higher than LLaMA’s and far above Gemma’s near-zero values, indicating model drift on domain-critical tokens.}
	\label{tab:olmo_frequent_elements}
\end{table*}

\paragraph{Why OLMo regressed while LLaMA and Gemma improved?}

We diagnose the effect of preference tuning by measuring forward
\(D_{\mathrm{KL}}(\text{DPO}\,\|\,\text{SFT})\) strictly on the \emph{generated continuation}:
we teacher–force the SFT decode, trim at EOS, and compute KL token‑wise. We also
summarize KL over a \emph{filtered token set} that carries the task semantics—element
symbols and multi‑digit numerals that encode compositions and phase fractions.
Under this lens, \textbf{LLaMA} shows small, localized KL bumps at decision bottlenecks;
\textbf{Gemma} remains close to its SFT policy; \textbf{OLMo} is different. Its KL spikes are
both larger and more frequent, and they land exactly on the filtered tokens. In
effect, the OLMo update reallocates probability mass on the symbols and numbers
that define alloy identity, not on harmless stylistic tokens (see Table~\ref{tab:olmo_frequent_elements}). This pattern
naturally explains the downstream regressions: if the largest distributional
shifts occur on element choices and volume proportions, the generator drifts
off the ``chemistry grammar'' that SFT had learned, degrading satisfaction of the
synthesis constraints.

\paragraph{Interpretation from the KL profiles}

\begin{figure*}[h]
	\centering
	\includegraphics[width=0.9\linewidth]{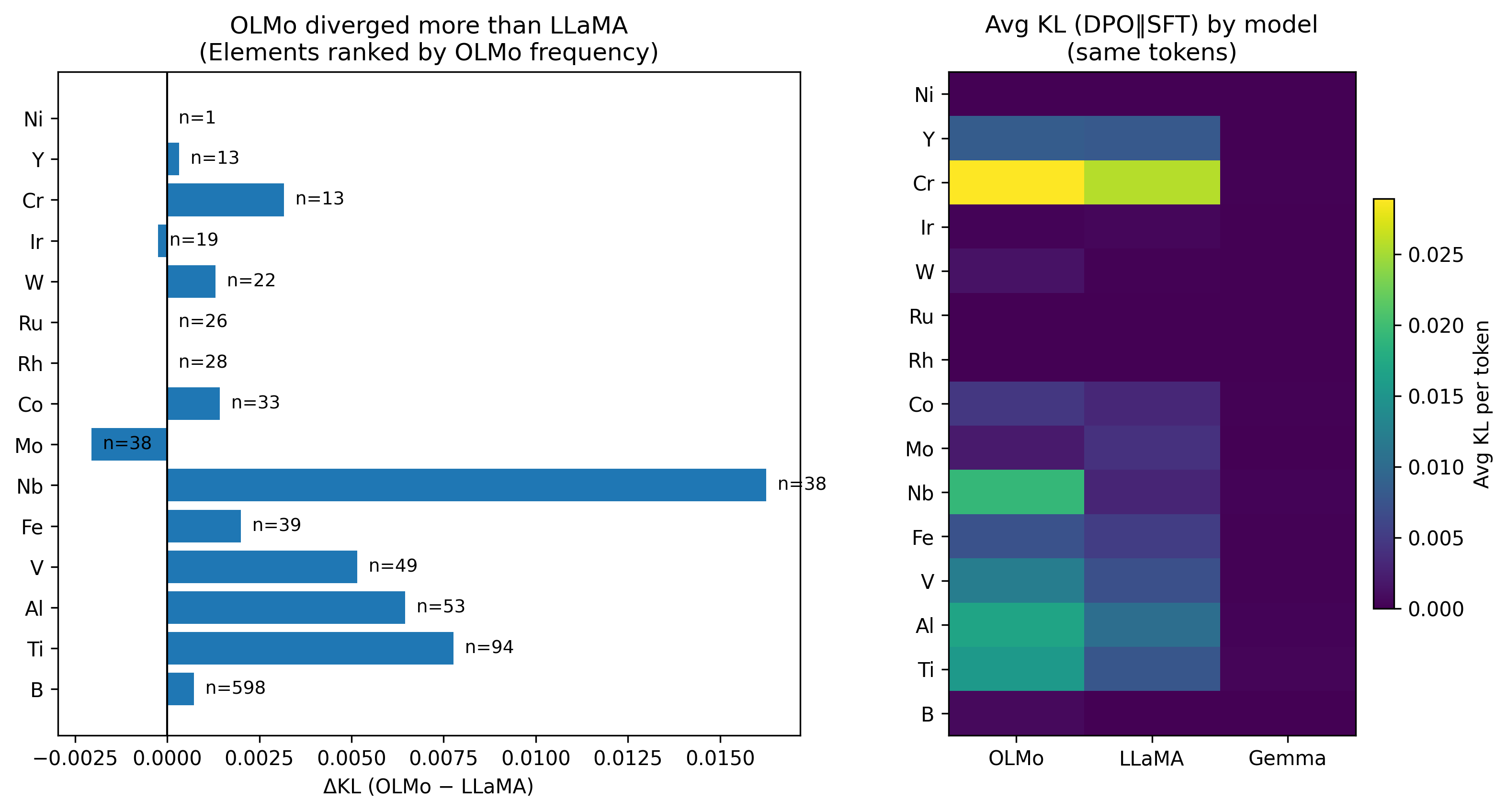}
	\caption{\textbf{OLMo goes out of distribution on domain‑critical element tokens after DPO.}
		\emph{Left:} Ranked bar chart of
		$\Delta\mathrm{KL}=\overline{D_{\mathrm{KL}}}(\mathrm{DPO}\,\|\,\mathrm{SFT})_{\text{OLMo}}
		-\overline{D_{\mathrm{KL}}}(\mathrm{DPO}\,\|\,\mathrm{SFT})_{\text{LLaMA}}$
		computed \emph{only on generated tokens} (teacher‑forced on the SFT continuation; trimmed at EOS).
		Elements are ordered by OLMo frequency; labels show OLMo occurrences ($n$).
		Positive bars indicate OLMo moved farther from its SFT reference than LLaMA did for the \emph{same} token.
		\emph{Right:} Heatmap of average per‑token $D_{\mathrm{KL}}(\mathrm{DPO}\,\|\,\mathrm{SFT})$ for the same elements across models (OLMo, LLaMA, Gemma).
		The consistently hotter OLMo column on key elements (e.g., Nb, Ti, Al, V) evidences over‑steer in the chemistry subspace where alloy identity is decided, while LLaMA shows moderate shifts and Gemma remains near the SFT policy.}
	\label{fig:deltaklbarplusheatmap}
\end{figure*}

The KL curves point to \emph{over‑steer} rather than lack of signal—a
strength–sensitivity mismatch between the DPO update and OLMo’s inductive bias.
(1) \emph{Architecture \(\times\) adapter placement/rank:} the same LoRA targets and rank that
are tame on LLaMA/Gemma appear to sit on more causal pathways in OLMo, so
identical gradients yield larger effective steps in logits for rare technical
tokens (elements, multi‑digit numerals). (2) \emph{Tokenizer/prior effects:} these tokens
live in a low‑frequency subspace; if OLMo’s pretraining allocates less robust
capacity there, the preference gradients induce higher variance and numeric
drift. (3) \emph{DPO hyperparameters:} a \(\beta\) and learning‑rate/step schedule that gently
nudges strong SFT policies (LLaMA/Gemma) can over‑correct a weaker or more
brittle SFT (OLMo), inflating KL precisely on the filtered token set. The net
effect is the signature we observe: the biggest divergence occurs where
correctness matters most (see Figure~\ref{fig:deltaklbarplusheatmap}).

\paragraph{Moving forward}
If we \emph{weaken and stabilize} the update in that subspace—e.g., increase
\(\beta\) (gentler preference step), reduce LR/steps or LoRA rank, and/or retarget
adapters (start with attention projections)—and optionally add a light reference
anchor (DPO‑KL or a small SFT CE mix‑in), the filtered‑token KL for OLMo should
drop into the LLaMA band. Under the same teacher‑forced evaluation, this KL
reduction should coincide with recovery on the synthesis objectives. In short,
the KL analysis localizes the failure mode (over‑steer on domain‑critical
tokens) and directly suggests how to fix it.

\end{document}